%% file: main.tex
\PassOptionsToPackage{dvipsnames,table}{xcolor}
\documentclass[letterpaper,twocolumn,10pt]{article}
\usepackage{usenix}
\usepackage{xcolor}

\input{includes/import}

\input{includes/solidity-highlighting}

\input{includes/command}

\begin{document}

\date{}
\title{\Large \bf Function Name Is All You Need to Detect Blockchain Application Attacks}

\input{includes/authors}

\maketitle

\begin{abstract}
Blockchain application attacks, targeting business logic bugs in decentralized applications (dApps), have been an increasing concern to their developers and users, causing significant financial loss.
Existing attack detectors either rely on handcrafted rules for detection, or need difficult-to-obtain smart contract source code to analyze attack transactions.
This makes them brittle and inapplicable in practice. 

In this paper, we argue that function name sequences suffice to capture the high-level semantics of a transaction, and hence can be used to detect blockchain application attacks.
Our empirical study on transactions from 424 real-world attack incidents shows that 98.46\% of call traces can be resolved to function names, whereas only 74.78\% invoke contracts with available source code.
Based on this observation, we propose \sysname{} (pronounced ``translucent''), an automated framework to detect blockchain application attacks by extracting application semantics from transaction call traces.
\sysname{} maps call traces to function name sequences and uses a transformer to learn semantics from such sequences.
Consequently, \sysname{} can detect attacks without relying on hand-coded patterns or source code.
Our results show that \sysname{} achieves a 1.56\% false negative rate on 424 known incidents with 14,611 attack transactions, and an estimated 0.0017\% false positive rate for benign transactions from over 500 million transactions on the Ethereum blockchain.
Finally, \sysname{} takes an average of 24.90 milliseconds to analyze a transaction, thus supporting real-time attack detection on popular blockchains.
\end{abstract}

\input{sections/01-intro.tex}
\input{sections/02-background.tex}
\input{sections/03-threat-model.tex}
\input{assets/fig-method-pipeline.tex}
\input{sections/04-challenge-insight.tex}
\input{sections/05-method.tex}
\input{sections/06-experiment.tex}
\input{sections/07-discussion.tex}
\input{sections/08-related-work.tex}
\input{sections/09-conclusion.tex}

\appendix

\input{appendices/01-ethics.tex}

\input{appendices/02-open-science.tex}

\bibliographystyle{plainurl}
\bibliography{ref}

\input{assets/table-attack-categories}
\input{appendices/03-solidity-evm-functions.tex}
\input{assets/table-appendix-comparison.tex}
\input{appendices/04-attack-categories.tex}
\input{appendices/05-function-name-long-tail.tex}
\input{appendices/06-baseline.tex}
\input{appendices/07-houston.tex}
\input{appendices/08-false-positives-analysis.tex}

\end{document}

%% file: includes/import.tex
\usepackage{amsmath}
\usepackage{amssymb}
\usepackage{amsthm}

\usepackage[inline]{enumitem}

\usepackage{graphicx}
\usepackage{subfig}
\usepackage{tikz}
\usetikzlibrary{tikzmark,calc}
\usetikzlibrary{positioning}
\usetikzlibrary{fit}
\usetikzlibrary{matrix}
\usetikzlibrary{arrows.meta}
\usetikzlibrary{decorations.text}
\usepackage{pgf-pie}

\usepackage{listings}
\usepackage{pifont}
\usepackage{booktabs}
\usepackage{float}

\usepackage{xstring}
\usepackage{xspace}

\hypersetup{breaklinks=true}
\usepackage{bookmark}

\makeatletter
\g@addto@macro{\UrlBreaks}{%
  \do\a\do\b\do\c\do\d\do\e\do\f\do\g\do\h\do\i\do\j%
  \do\k\do\l\do\m\do\n\do\o\do\p\do\q\do\r\do\s\do\t%
  \do\u\do\v\do\w\do\x\do\y\do\z%
  \do\A\do\B\do\C\do\D\do\E\do\F\do\G\do\H\do\I\do\J%
  \do\K\do\L\do\M\do\N\do\O\do\P\do\Q\do\R\do\S\do\T%
  \do\U\do\V\do\W\do\X\do\Y\do\Z%
  \do\0\do\1\do\2\do\3\do\4\do\5\do\6\do\7\do\8\do\9}
\makeatother

%% file: includes/solidity-highlighting.tex
\lstdefinelanguage{Solidity}{
    keywords=[1]{anonymous, assembly, assert, balance, break, call, callcode, case, catch, class, constant, continue, constructor, contract, debugger, default, delegatecall, delete, do, else, emit, event, experimental, export, external, false, finally, for, function, gas, if, implements, import, in, indexed, instanceof, interface, internal, is, length, library, log0, log1, log2, log3, log4, solidity, storage, struct, suicide, super, switch, then, this, throw, transfer, true, try, typeof, using, value, view, while, addmod, ecrecover, keccak256, mulmod, ripemd160, sha256, sha3},
    keywordstyle=[1]\color{blue}\bfseries,
    keywords=[2]{address, bool, byte, bytes, bytes1, bytes2, bytes3, bytes4, bytes5, bytes6, bytes7, bytes8, bytes9, bytes10, bytes11, bytes12, bytes13, bytes14, bytes15, bytes16, bytes17, bytes18, bytes19, bytes20, bytes21, bytes22, bytes23, bytes24, bytes25, bytes26, bytes27, bytes28, bytes29, bytes30, bytes31, bytes32, enum, int, int8, int16, int24, int32, int40, int48, int56, int64, int72, int80, int88, int96, int104, int112, int120, int128, int136, int144, int152, int160, int168, int176, int184, int192, int200, int208, int216, int224, int232, int240, int248, int256, mapping, string, uint, uint8, uint16, uint24, uint32, uint40, uint48, uint56, uint64, uint72, uint80, uint88, uint96, uint104, uint112, uint120, uint128, uint136, uint144, uint152, uint160, uint168, uint176, uint184, uint192, uint200, uint208, uint216, uint224, uint232, uint240, uint248, uint256, var, void, ether, finney, szabo, wei, days, hours, minutes, seconds, weeks, years},
    keywordstyle=[2]\color{teal}\bfseries,
    keywords=[3]{block, blockhash, coinbase, difficulty, gaslimit, number, timestamp, msg, data, gas, sender, sig, value, now, tx, gasprice, origin},
    keywordstyle=[3]\color{violet}\bfseries,
    identifierstyle=\color{black},
    sensitive=true,
    comment=[l]{//},
    morecomment=[s]{/*}{*/},
    commentstyle=\color{gray}\ttfamily,
    stringstyle=\color{red}\ttfamily,
    morestring=[b]',
    morestring=[b]"
}

%% file: includes/command.tex
\newcommand{\sysname}{TxLucent}
\newcommand{\cream}{Cream incident}
\newcommand{\etal}{\textit{et al}.\xspace}
\newcommand{\ie}{\textit{i}.\textit{e}.,\xspace}
\newcommand{\eg}{\textit{e}.\textit{g}.,\xspace}

\newcommand{\arrow}{\ensuremath{\rightarrow}}

\renewcommand{\texttt}[1]{%
  \begingroup
    \ttfamily
    \StrSubstitute{#1}{.}{.\allowbreak}[\tempStr]%
    \tempStr%
  \endgroup
}

\newcommand{\insight}[2]{%
  \begin{quote}
    \textbf{#1}~#2
  \end{quote}
}

\newenvironment{codebox}{}{}

%% file: includes/authors.tex
\author{
    {\rm Rui Xi}\\
    University of British Columbia
    \and
    {\rm Zehua Wang}\\
    University of British Columbia
    \and
    {\rm Karthik Pattabiraman}\\
    University of British Columbia
}

\hypersetup{
    pdftitle={Function Name Is All You Need to Detect Blockchain Application Attacks},
    pdfauthor={Rui Xi; Zehua Wang; Karthik Pattabiraman}
}

%% file: sections/01-intro.tex
\section{Introduction}
\label{sec:introduction}

Decentralized applications (dApps) are built on top of smart contracts, which are self-executing programs on blockchains, enabling a wide array of functionalities, \eg{} issuing new cryptocurrencies.
However, as the adoption of dApps has surged, so have security attacks on them.
The most significant are \emph{blockchain application attacks}, which arise from business logic bugs in dApps~\cite{zhang2023demystifying,zhou2023sok} and have led to financial losses amounting to billions of dollars~\cite{defillama-hack}.
These attacks have constituted 42\% of all incidents on popular blockchains since 2022~\cite{zhou2023sok}, and their number continues to grow~\cite{DeFiHackLabs}.
They typically occur when one dApp (A) interacts with another dApp (B) whose business logic deviates from A's intended behavior.
In such cases, an attacker may manipulate B's state to compromise A, thereby gaining a financial advantage.

Existing studies use different strategies to detect blockchain application attacks.
There are two broad approaches based on the features they utilize, \ie off-chain and on-chain.
Wu \etal~\cite{wu2025hunting}, hereafter referred to as ``Hunting'' (based on its paper title), use smart contract source code and comments to infer expected dApp semantics, and  detect attacks by checking whether a transaction deviates from these semantics.
However, relying on such \emph{off-chain features}, \ie{} source code and comments, makes it difficult to generalize. 	In our experiments, we found that only 74.78\% of call traces invoke a smart contract whose source code is available  (Section~\ref{sec:RQ1}).

In contrast, CLUE~\cite{qin2025enhancing} leverages \emph{on-chain features}, \ie features that are readily available on the blockchain.
It lifts low-level features into intermediate representations and uses patterns to detect attacks.
However, significant manual effort and domain knowledge are needed to craft the patterns for detection.
Moreover, the continued evolution of dApps requires ongoing manual analysis, making this a never-ending task.

\emph{In this paper, we posit that function name sequences are an effective way to detect application-level attacks on blockchain applications. }
Our hypothesis is based on the following two observations: function names are
\begin{enumerate*}[label=(\roman*)]
\item \textbf{semantically rich} because smart contract developers typically adopt descriptive function names, which are often more informative than low-level features; and
\item \textbf{more widely available than source code}. 
This is because source code lookup is contract address-level, whereas function name lookup is function selector-level.
Because the same selector can recur across contracts, a name extracted from one contract can map calls to other contracts whose source code is unavailable. 
Function name sequences can therefore cover more call traces than source-level features. %
\end{enumerate*}

We propose \sysname{}, a framework for detecting blockchain application attacks using function name sequences.
We call it \sysname{} (pronounced ``translucent'') because function names alone provide an informative but incomplete view of transaction semantics---a translucent view that our results show is nevertheless effective for attack detection.
\sysname{} has two steps.
First, given a transaction, \sysname{} extracts functions from the transaction's call traces and maps them to a sequence of function names.
Second, it uses a classifier to learn the semantic characteristics of attacks for effective attack detection.
We choose a Transformer to leverage its attention mechanism and capture semantics in both attack and benign transactions~\cite{vaswani2017attention}.
Our key contribution is not the choice of the classifier model, but rather the demonstration that function name sequences suffice to distinguish attack transactions from benign ones without handcrafted rules, contract source code, or heavyweight feature reconstruction.

\emph{To the best of our knowledge, \sysname{} is the first technique to extract transaction semantics from function name sequences and use them for blockchain application attack detection.}

\textbf{Contributions}: In summary, our contributions are:
\begin{itemize}
    \item We observe that function name sequences from call traces %
    can effectively detect attacks without requiring either source code or on-chain feature engineering. 
    \item We propose \sysname{}, a two-step detector for blockchain application attacks.
    First, \sysname{} maps call traces to function name sequences via the Sequence builder.
    Second, it encodes the sequences and uses a Transformer classifier to separate attacks from benign transactions.
    \item We perform an empirical study to analyze what proportion of real-world call traces can be covered by function name sequences versus smart contracts' source code. %
    
    \item We design an attack detector evaluation workflow based on real-world attacks on blockchain applications.
    We construct a set of 424 incidents from DeFiHackLab, a community-maintained open-source attack database~\cite{DeFiHackLabs} with 14,611 unique attack transactions, to train \sysname{}.
    We evaluate \sysname{} on the DeFiHackLab dataset (excluding the training set) to measure False Negative Rate (FNR), and on the Ethereum benign dataset consisting of 537,069,442 benign transactions from Ethereum to measure False Positive Rate (FPR).
\end{itemize}

\textbf{Results}: (1) In our datasets, function-name coverage is 98.46\%, while only 74.78\% of the calls invoke contracts with source code.
(2) During training, we apply incident-level five-fold cross-validation to avoid incident leakage.
Each checkpoint receives one fold for evaluation, and only the other four folds for training.
Across all checkpoints, \sysname{} achieves an FNR of 1.56\%.
On the same incidents, CLUE and Hunting obtain FNR values of 70.77\% and 83.17\%, respectively.
(3) \sysname{} has an FPR of 0.0017\% on the Ethereum benign dataset.
(4) \sysname{} incurs an average detection time of 24.90 milliseconds per transaction, supporting realtime detection.
(5) \sysname{}'s detection results agree with their official incident reports with a median alignment score of 74.86\%.
(6) Even when attackers attempt to proactively evade \sysname{} %
its FNR increases only marginally from 1.56\% to 2.26\%.

%% file: sections/02-background.tex
\section{Background}
\label{sec:background}

\subsection{Ethereum and Smart Contracts}
\label{sec:background:ethereum}

Blockchains enable decentralized, transparent record-keeping across nodes~\cite{nakamoto2008bitcoin}.
Unlike centralized systems, blockchains remove intermediaries while preserving data integrity.
Ethereum~\cite{buterin2013ethereum} extends this with smart contracts.
Its success spurred other EVM-compatible (Ethereum Virtual Machine) blockchains, fostering a decentralized ecosystem.

Smart contracts on EVM-compatible blockchains have their compiled bytecode stored \emph{on-chain}.
Solidity is the dominant high-level language for writing such contracts.
We summarize how Solidity and the EVM handle functions and how this information appears in transaction call traces in Appendix~\ref{sec:appendix:background:function}.
Developers often publish Application Binary Interfaces (ABIs) to improve smart-contract transparency and interoperability.
An ABI, akin to an Application Programming Interface (API), defines how applications exchange data.
For example, the ERC20 standard exposes an interface (commonly \texttt{IERC20}) with methods such as \texttt{transfer(address,uint256)}; a dApp can invoke a token transfer via \texttt{IERC20(token).transfer(to, amount)}.
Both source code and ABIs are \emph{off-chain},
but what differentiates an ABI from source code is that an ABI defines only the function signatures, not contract logic or implementation details, and is often available through ERC specifications and APIs~\cite{eips,blockscout-contract-abi-api,sourcify-contract-api}.
The community also developed de facto standards for ABIs in specific dApp categories.

\subsection{dApps and Ethereum Request for Comments (ERCs)}
\label{sec:background:dapps}

dApps consist of a set of smart contracts that work together to serve a specific use case.
Ethereum Request for Comments (ERCs) are community standards defining common dApp interfaces.
ERCs specify function signatures (\ie function names, parameter types, and return types) that any conforming dApp must implement.
This standardization lets any dApp that understands the ERC interface interact with any ERC-compliant token without knowing its implementation.

Tokens are among the most fundamental dApp categories, representing digital assets that can encapsulate utilities, securities, or votes.
ERC20~\cite{erc20}, the most widely adopted token standard, defines function signatures that any ERC20-compliant token must implement, such as \texttt{transfer}, \texttt{approve}, and \texttt{transferFrom}.
Any token implementing ERC20 exposes these exact function names.
As a result, thousands of tokens on EVM-compatible blockchains share function names for the same operations.
Similarly, ERC721~\cite{erc721} standardizes non-fungible token operations such as \texttt{ownerOf}.

Decentralized exchanges (DEXs) are tokenization-based dApps that enable token trading, often via automated market makers (AMMs)~\cite{automated-market-maker}.
While DEX interfaces are not standardized as ERCs, popular AMM designs converge on de facto method names for trading and liquidity management.
For example, the Uniswap V2 router interface~\cite{uniswap,uniswap-v2-periphery} exposes \texttt{swapExactTokensForTokens}, \texttt{addLiquidity}, and \texttt{removeLiquidity}, while Curve pools~\cite{Curve} provide analogous entry points such as \texttt{exchange}.
DEXs therefore reuse function names for the same operations.

Lending platforms, introduced in Section~\ref{sec:introduction}, are dApps for borrowing and lending.
ERC4626~\cite{erc4626} is a widely used interface standard for tokenized vaults, defining exact function names such as \texttt{deposit}, \texttt{withdraw}, \texttt{mint}, and \texttt{redeem}.
In addition, ERC3156~\cite{erc3156} standardizes single-asset flash loans via \texttt{flashLoan} and the borrower callback \texttt{onFlashLoan}.
Beyond these standardized interfaces, conventions in lending platforms include \texttt{supply}, \texttt{borrow}, \texttt{repay}, and \texttt{liquidate}.

Because formal ERCs and de facto conventions encourage consistent naming across dApps, function names can carry rich semantic information about business operations.
We next illustrate these concepts with the \cream.

\section{Motivating Example}
\label{sec:background:motivating}

\input{assets/fig-cream-illus.tex}

Figure~\ref{fig:cream-finance-illustration} shows the workflow of the lending dApp C.R.E.A.M. Finance~\cite{cream-finance} and a real-world blockchain application attack on it~\cite{cream-finance-blog}, which we refer to as the \cream{} (we use this as a running example).
This lending dApp allows users to \ding{172} borrow assets based on the amount of collateral or deposited assets, establishing a ``borrow room'' that limits the total debt~\cite{compound2019money,aave2020protocol}.
Once fully utilized, borrowing is disallowed until the borrower \ding{173} repays the existing debt and \ding{175} reclaims their collateral.
If asset prices fall, \ding{174} liquidation is triggered when collateral drops below a threshold relative to debt.
In liquidation, any user may repay at a discount and \ding{175} seize the collateral.
The borrower bears the discount loss.
Intuitively, we can use a sequence of operations to represent these workflows, \eg{} \texttt{borrow} $\rightarrow$ \texttt{repay} $\rightarrow$ \texttt{seize} and \texttt{borrow} $\rightarrow$ \texttt{liquidate} $\rightarrow$ \texttt{seize}.

In the \cream{}, the attacker exploited a reentrancy vulnerability in the lending dApp by first maxing out the borrow room through an initial \ding{176} borrowing of AMP tokens.
Subsequently, the attacker reentered the contract to bypass the borrow limit checks and executed an additional \ding{177} borrowing.
This reentrant borrowing led to a situation where the collateral's value no longer adequately covered the debt.
Consequently, the attacker \ding{178} voluntarily initiated liquidation to \ding{179} seize their collateral.
Despite incurring losses from liquidation, the cumulative debt generated by the reentrant borrowings exceeded the liquidation cost and discount loss, thus financially benefiting the attacker.
Similarly, we can represent the attack transaction as \texttt{borrow} $\rightarrow$ \texttt{borrow} $\rightarrow$ \texttt{liquidate} $\rightarrow$ \texttt{seize}.
The attack occurs at and can be identified by the second \texttt{borrow} that occurs without an intervening \texttt{repay} to reduce the position.
This example reveals the intuition behind our work: 
\emph{the function name sequence represents the high-level business semantics of a transaction, and can be used to detect blockchain application attacks.}

We examine the on-chain features of the \cream{} attack transaction and the off-chain features of the Cream and underlying dApps.
Listing~\ref{listing:cream-finance-code} shows the contracts involved and their function signatures.

\input{assets/list-cream-code.tex}

Five contracts are involved in this attack: the victim lending contracts \texttt{AMPMarket} and \texttt{ETHMarket};
the collateral storage \texttt{DApp}; the borrowed asset \texttt{AMPToken} (whose \texttt{transferFrom} triggers the \texttt{tokensReceived} callback, enabling reentrancy);
and the attacker-controlled \texttt{Attack} contract. The attacker supplies collateral via \texttt{AMPMarket.supply}, then calls \texttt{AMPMarket.borrow}.
When \texttt{AMPMarket.borrow} internally calls \texttt{AMPToken.transferFrom} to transfer the borrowed tokens, \texttt{AMPToken} invokes \texttt{Attack.tokensReceived}, which reenters via \texttt{ETHMarket.borrow} before the debt state from the first borrow is updated.
This allows borrowing beyond the permitted collateral limit.
The attacker then calls \texttt{AMPMarket.liquidate} to seize collateral from other users.

The root cause is that the lending dApp fails to account for the \texttt{AMPToken.tokensReceived} callback, which may reenter \texttt{ETHMarket.borrow}, leading to reentrancy~\cite{xue2020cross} across \texttt{ETHMarket} and \texttt{AMPMarket}.
The call traces in Listing~\ref{lst:call-trace-tree} show how the reentrant \texttt{borrow} call subverts the expected \texttt{borrow} $\rightarrow$ \texttt{liquidate} $\rightarrow$ \texttt{seize} lifecycle.
For clarity, we show function names rather than selectors.

We then analyze the availability of these features for the \cream.

\emph{On-chain features.} The attack transaction call traces are publicly available on the blockchain once the transaction is submitted to the public mempool or mined into a block.
Anyone can obtain call traces through a node provider or their own full blockchain node; blockchain explorers such as Etherscan expose related internal-call data through documented APIs~\cite{etherscan-internal-transactions-api}.

\emph{Off-chain features.} The function signatures of the involved contracts are not on-chain but are available off-chain:
The \texttt{AMPToken} contract follows the ERC20 standard, whose ABI is publicly available.
The \texttt{AMPMarket} and \texttt{ETHMarket} contracts are part of the Cream protocol, whose ABIs are also publicly available.
The \texttt{Attack} contract is attacker-controlled, so its source code and ABI are not publicly available; however, its \texttt{tokensReceived} function signature is available because it is standardized by ERC777.
Hence, all function signatures are off-chain except the attacker's \texttt{start} function.

%% file: assets/fig-cream-illus.tex
\begin{figure}[!t]
    \centering
    \begin{minipage}{0.45\textwidth}
        \centering
        \begin{tikzpicture}[
            >=Stealth,
            every node/.style={font=\footnotesize},
            box/.style={draw, rectangle, line width=0.55pt, minimum height=0.42cm, inner sep=1.5pt, align=center},
            flow/.style={->, line width=0.55pt},
            lab/.style={font=\footnotesize, fill=white, inner sep=0.5pt}
        ]
            \node (creamA) at (0,0) [box, minimum width=3.2cm] {C.R.E.A.M. Finance};
            \node (ethA)   at (0,-0.42) [box, minimum width=3.2cm] {ETH market};
            \node (borrower) at (-2.45,-1.85) [box, minimum width=2.2cm] {Borrower};
            \node (liquidator) at (2.45,-1.85) [box, minimum width=2.2cm] {Liquidator};
            
            \node[anchor=north west] at (-4,0.5) {(a)};
            
            \draw[flow, bend left=18] (ethA) to node[midway, lab] {\ding{172}} (borrower);
            \draw[flow, bend left=18] (borrower) to node[midway, lab] {\ding{173}} (ethA);
            \draw[flow, bend left=18] (liquidator) to node[midway, lab] {\ding{174}} (ethA);
            \draw[flow, bend left=18] (ethA) to node[midway, lab] {\ding{175}} (liquidator);
        \end{tikzpicture}
    \end{minipage}
    \hfill
    \begin{minipage}{0.45\textwidth}
        \centering
        \begin{tikzpicture}[
            >=Stealth,
            every node/.style={font=\footnotesize},
            box/.style={draw, rectangle, line width=0.55pt, minimum height=0.42cm, inner sep=1.5pt, align=center},
            flow/.style={->, line width=0.55pt},
            lab/.style={font=\footnotesize, fill=white, inner sep=0.5pt}
        ]
            \node (creamB) at (0,0) [box, minimum width=3.8cm] {C.R.E.A.M. Finance};
            \node (ampmarket) at (-0.95,-0.42) [box, minimum width=1.9cm] {AMP market};
            \node (ethmarket) at (0.95,-0.42) [box, minimum width=1.9cm] {ETH market};
            \node (attackBorrower) at (-2.45,-1.85) [box, minimum width=2.2cm] {Attacker 1};
            \node (attackLiquidator) at (2.45,-1.85) [box, minimum width=2.2cm] {Attacker 2};
            
            \node[anchor=north west] at (-4,0.5) {(b)};
            
            \draw[flow, bend left=18] (attackBorrower) to node[midway, lab] {\ding{176}} (ampmarket);
            \draw[flow, bend right=18] (attackBorrower) to node[midway, lab] {\ding{177}} (ethmarket);
            \draw[flow, bend right=18] (attackLiquidator) to node[midway, lab] {\ding{178}} (ampmarket);
            \draw[flow, bend right=18] (ampmarket) to node[midway, lab] {\ding{179}} (attackLiquidator);
        \end{tikzpicture}
    \end{minipage}
    \caption{Illustration of C.R.E.A.M. Finance in the (a) regular and (b) attack transactions.}
    \label{fig:cream-finance-illustration}
\end{figure}
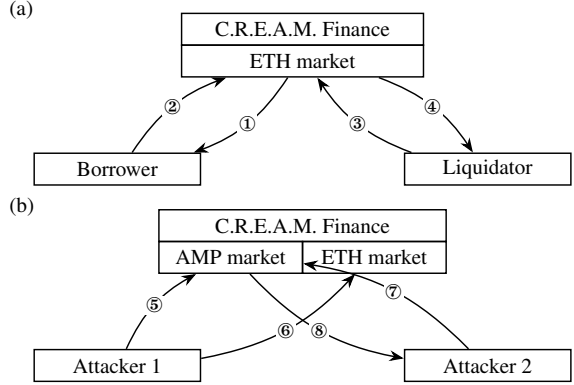

%% file: assets/list-cream-code.tex
\begin{figure}[!t]
	\begin{footnotesize}
	\begin{codebox}
	\begin{lstlisting}[language=Solidity,escapeinside={§}{§}]
contract DApp {
	mapping(address => uint) collaterals; }
contract §\texttt{AMPToken}§ {
	function transferFrom(address sender, address receiver, uint amount){...}}
contract AMPMarket {
	function supply(uint amount) public {...}
	function borrow(uint amount) public {...}
	function liquidate(address borrower) public {...}
	function _repay(address borrower, uint amount) {...}}
contract ETHMarket {
	function supply(uint amount) public {...}
	function borrow(uint amount) public {...}}
contract Attack {
	function start() {...}
	function tokensReceived() {...}}
\end{lstlisting}
	\end{codebox}
	\captionsetup{name=Listing}
	\caption{Simplified contract interfaces for the \cream{}.}
	\label{listing:cream-finance-code}
	\end{footnotesize}
\end{figure}

%% file: sections/03-threat-model.tex
\section{Threat Model, Scope and Assumptions}
\label{sec:threat-model}

We consider a blockchain system with blockchain infrastructure, dApps implemented as smart contracts, users interacting with dApps via transactions, and an attacker (a malicious user) seeking to exploit dApp vulnerabilities by crafting abnormal transactions.

Our scope is blockchain application attacks on EVM-compatible blockchains.
We exclude attacks on the underlying blockchain infrastructure (\eg 51\% attacks~\cite{nakamoto2008bitcoin} and eclipse attacks~\cite{heilman2015eclipse}).
We also exclude attacks targeting users (\eg phishing and private-key theft), Solidity language-feature issues (\eg integer overflow and precision loss), and insider attacks.
The attacker has standard user capabilities to submit transactions to dApps and may deploy auxiliary smart contracts to automate the attack~\cite{liu2025reentrancy}.
The attack surface is business-logic flaws exploitable through smart contract-defined interactions to manipulate dApp state or behavior; the attacker's objectives are financial gain or disruption of dApp functionality.
The attacker is assumed to know the dApp's business logic and source code, analyze the contract code and documentation to identify vulnerabilities, and use prior attack patterns and techniques to craft malicious transactions~\cite{zhou2023sok}.

We further assume users have access to on-chain features, \ie{} transaction metadata and call traces, without network latency, as they can be obtained from a local full blockchain node.
We assume that the source code of a contract appearing in a transaction is \emph{not} available. Instead, \sysname{} uses a selector-to-name database built once from public sources; selectors absent from this database map to \texttt{[UNK\_SELECTOR]}.
Section~\ref{sec:RQ1} empirically measures the resulting name coverage, including calls to contracts with their source code available.

Our goal is accurate, timely detection of blockchain application attacks on EVM-compatible blockchains.
Specifically, we target low false positive and false negative rates and real-time detection, so attacks can be identified and mitigated before causing further damage.

%% file: assets/fig-method-pipeline.tex
\begin{figure*}[!t]
  \centering

  \begingroup
  \usetikzlibrary{arrows.meta, positioning, calc, shapes, backgrounds}

  \newcommand{\txlucentDrawFullNode}[1]{%
    \begin{scope}[shift={(#1)}, scale=0.048, thick]
      \begin{scope}[shift={(-12,-12)}]
        \draw (2,22) rectangle (22,16); \fill (5,19) circle (1); \draw (8,19) -- (19,19);
        \draw (2,15) rectangle (22,9);  \fill (5,12) circle (1); \draw (8,12) -- (19,12);
        \draw (2,8)  rectangle (22,2);  \fill (5,5)  circle (1); \draw (8,5)  -- (19,5);
      \end{scope}
    \end{scope}
  }

  \newcommand{\txlucentDrawCallTraces}[1]{%
    \begin{scope}[shift={(#1)}, scale=0.048, thick]
      \begin{scope}[shift={(-12,-12)}]
        \fill (8,22) rectangle (24,20);
        \draw (4,21) -- (8,21);
        \draw (4,21) -- (4,5);

        \draw (4,15) -- (8,15);
        \fill (8,16) rectangle (20,14);

        \draw (6,15) -- (6,9) -- (8,9);
        \fill (8,10) rectangle (16,8);

        \draw (4,5) -- (8,5);
        \fill (8,6) rectangle (18,4);
      \end{scope}
    \end{scope}
  }

  \newcommand{\txlucentDrawFNDB}[1]{%
    \begin{scope}[shift={(#1)}, scale=0.04, thick]
      \begin{scope}[shift={(-12,-12)}]
        \draw (12,22) ellipse (8 and 3);
        \draw (4,22) -- (4,2);
        \draw (20,22) -- (20,2);
        \draw (4,2) arc (180:360:8 and 3);
        \draw[gray] (4,15) arc (180:360:8 and 3);
        \draw[gray] (4,8)  arc (180:360:8 and 3);
      \end{scope}
    \end{scope}
  }

  \newcommand{\txlucentDrawPosition}[1]{%
    \begin{scope}[shift={(#1)}, scale=0.04, thick]
      \begin{scope}[shift={(-12,-12)}]
        \draw (12,22) -- (7,12);  \draw (12,22) -- (17,12);
        \draw (7,12) -- (4,2);    \draw (7,12) -- (10,2);
        \draw (17,12) -- (14,2);  \draw (17,12) -- (20,2);

        \fill[white] (12,22) circle (2.5); \draw (12,22) circle (2.5);
        \fill[white] (7,12)  circle (2.5); \draw (7,12)  circle (2.5);
        \fill[white] (17,12) circle (2.5); \draw (17,12) circle (2.5);

        \fill[white] (4,2)  circle (1.5); \draw (4,2)  circle (1.5);
        \fill[white] (10,2) circle (1.5); \draw (10,2) circle (1.5);
        \fill[white] (14,2) circle (1.5); \draw (14,2) circle (1.5);
        \fill[white] (20,2) circle (1.5); \draw (20,2) circle (1.5);
      \end{scope}
    \end{scope}
  }

  \newcommand{\txlucentDrawFName}[1]{%
    \begin{scope}[shift={(#1)}, scale=0.04, thick]
      \begin{scope}[shift={(-12,-12)}]
        \draw (0,24) rectangle (24,0);
        \node at (12,12) {\footnotesize $f(x)$};
      \end{scope}
    \end{scope}
  }

  \newcommand{\txlucentDrawPosEnc}[1]{%
    \begin{scope}[shift={(#1)}, scale=0.04, thick]
      \begin{scope}[shift={(-12,-12)}]
        \draw (0,24) rectangle (24,0);
        \draw[domain=2:22, samples=50, smooth, thick] plot (\x, {12 + 8*sin((\x-2)*360/20)});
      \end{scope}
    \end{scope}
  }

  \newcommand{\txlucentDrawEmbeddings}[1]{%
    \begin{scope}[shift={(#1)}, scale=0.04, thick]
      \begin{scope}[shift={(-12,-12)}]
        \draw[thick] (6,24) -- (2,24) -- (2,0) -- (6,0);
        \draw[thick] (18,24) -- (22,24) -- (22,0) -- (18,0);
        \fill (8,20) rectangle (10,4);
        \fill (11,20) rectangle (13,4);
        \fill (14,20) rectangle (16,4);
      \end{scope}
    \end{scope}
  }

  \newcommand{\txlucentDrawTransformer}[1]{%
    \begin{scope}[shift={(#1)}, scale=0.048, thick]
      \begin{scope}[shift={(-12,-12)}]
        \draw[dashed] (0,24) rectangle (24,0);
        \foreach \y in {4, 12, 20} {
          \foreach \yy in {4, 12, 20} {
            \draw[gray, thin] (6, \y) -- (18, \yy);
          }
        }
        \foreach \y in {4, 12, 20} {
          \fill[black] (6, \y) circle (2.5);
          \fill[white] (18, \y) circle (2.5);
          \draw (18, \y) circle (2.5);
        }
      \end{scope}
    \end{scope}
  }

  \newcommand{\txlucentDrawLinear}[1]{%
    \begin{scope}[shift={(#1)}, scale=0.048, thick]
      \begin{scope}[shift={(-12,-12)}]
        \draw (12,12) circle (11);
        \draw (1,12) -- (3,12); \draw (21,12) -- (23,12);
        \draw (12,1) -- (12,3); \draw (12,21) -- (12,23);
        \draw[very thick] (5,8) -- (12,8) -- (17,16);
      \end{scope}
    \end{scope}
  }

  \newcommand{\txlucentDrawResult}[1]{%
    \begin{scope}[shift={(#1)}, scale=0.048, thick]
      \begin{scope}[shift={(-8,-12)}]
        \draw[thick] (-2,12) -- (8,12);
        \draw[thick] (8,12) -- (16,22);
        \draw[thick] (8,12) -- (16,2);
        \fill (16,22) circle (2);
        \fill (16,2) circle (2);
        \node[anchor=west, font=\footnotesize, inner sep=1pt] at (16.5,22) {Attack};
        \node[anchor=west, font=\footnotesize, inner sep=1pt] at (16.5,2) {Benign};
      \end{scope}
    \end{scope}
  }

  \begin{tikzpicture}[
    >=Stealth,
    lbl/.style={align=center, font=\footnotesize, text=black},
    every label/.style={align=center, font=\footnotesize, text=black},
    pipe/.style={->, line width=0.55pt},
    group/.style={draw, dashed, line width=0.55pt, fill=black!2, inner sep=0pt}
  ]

    \node[group, minimum width=7.25cm, minimum height=4.15cm,
      anchor=south west] (seqbox) at (-0.25,-2.00) {};
    \node[anchor=north west, font=\footnotesize\bfseries] at (-0.05,1.90) {Sequence builder};

    \node[group, minimum width=9.30cm, minimum height=4.15cm,
      anchor=south west] (clfbox) at (7.45,-2.00) {};
    \node[anchor=north west, font=\footnotesize\bfseries] at (7.65,1.90) {Classifier};

    \node[lbl, label=below:{Full Node\\Provider}, minimum width=1.08cm, minimum height=1.08cm] (fullnode) at (0.75, 0.00) {};
    \txlucentDrawFullNode{fullnode.center}

    \node[lbl, label=below:{Call Traces}, minimum width=1.08cm, minimum height=1.08cm] (calltraces) at (2.55, 0.00) {};
    \txlucentDrawCallTraces{calltraces.center}

    \node[lbl, label=below:{Position}, minimum width=0.90cm, minimum height=0.90cm] (position) at (4.90, 0.92) {};
    \txlucentDrawPosition{position.center}

    \node[lbl, label=below:{Selector-to-Name\\Database}, minimum width=0.90cm, minimum height=0.90cm] (fndb) at (4.35, -0.76) {};
    \txlucentDrawFNDB{fndb.center}

    \node[lbl, label=below:{Function\\Name}, minimum width=0.90cm, minimum height=0.90cm] (fname) at (6.15, -0.76) {};
    \txlucentDrawFName{fname.center}

    \node[lbl, label=below:{Positional\\Encoding}, minimum width=0.90cm, minimum height=0.90cm] (posenc) at (8.70, 0.92) {};
    \txlucentDrawPosEnc{posenc.center}

    \node[lbl, label=below:{Embeddings}, minimum width=0.90cm, minimum height=0.90cm] (embeddings) at (8.70, -0.76) {};
    \txlucentDrawEmbeddings{embeddings.center}

    \node[lbl, label=below:{Transformer}, minimum width=1.08cm, minimum height=1.08cm] (transformer) at (11.15, 0.00) {};
    \txlucentDrawTransformer{transformer.center}

    \node[lbl, label=below:{Linear}, minimum width=1.08cm, minimum height=1.08cm] (linear) at (13.35, 0.00) {};
    \txlucentDrawLinear{linear.center}

    \node[lbl, label=below:{Result}, minimum width=2.20cm, minimum height=1.08cm] (result) at (15.00, 0.00) {};
    \txlucentDrawResult{result.center}

    \draw[pipe] (fullnode.east) -- (calltraces.west);

    \draw[pipe] (calltraces.east) to[out=0, in=180] (position.west);
    \draw[pipe] (calltraces.east) to[out=0, in=180] (fndb.west);

    \draw[pipe] (fndb.east) -- (fname.west);

    \draw[pipe] (position.east) -- (posenc.west);
    \draw[pipe] (fname.east) -- (embeddings.west);

    \draw[pipe] (posenc.east) to[out=0, in=180] (transformer.west);
    \draw[pipe] (embeddings.east) to[out=0, in=180] (transformer.west);

    \draw[pipe] (transformer.east) -- (linear.west);
    \draw[pipe] (linear.east) -- (result.west);

  \end{tikzpicture}
  \endgroup

  \caption{High-level architecture of \sysname{}. Sequence builder reconstructs call traces to function names; Classifier takes function name and positional features to detect attacks.}
  \label{fig:method-pipeline}
\end{figure*}
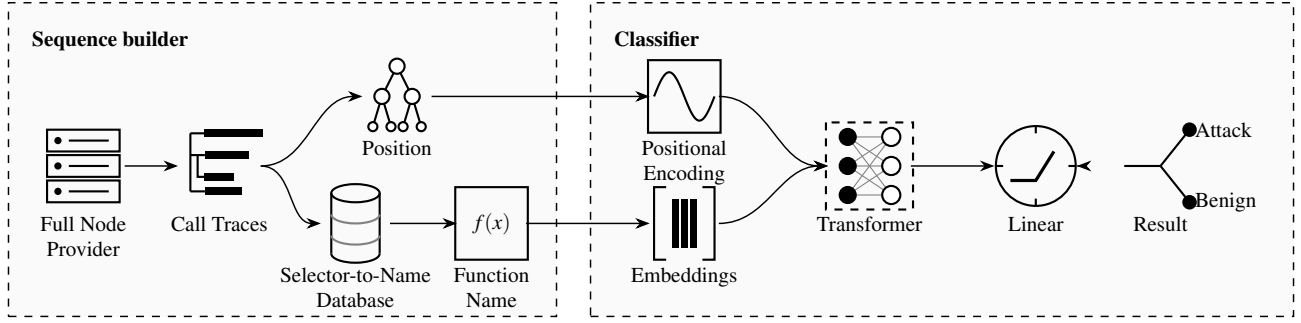

%% file: sections/04-challenge-insight.tex
\section{Challenges and Insights}
\label{sec:challenge-insight}

\subsection{Challenges}
\label{sec:challenge-insight:challenges}

Although blockchain data are transparent and traceable, detecting blockchain application attacks remains challenging: on-chain features are available but low-level, while off-chain features are semantically richer but harder to obtain.

\textbf{Challenge 1: On-chain features.}
Blockchain transactions expose abundant low-level on-chain features, but detecting blockchain application attacks requires high-level semantics.

As discussed in Section~\ref{sec:background:ethereum}, a transaction call trace records the sender, receiver, and function selector.
Other on-chain features include logs, transaction receipts, and state changes.
Attack intent emerges only from the combination of these artifacts.
As of 2025, Ethereum had more than 4 million deployed contracts and 360 million unique addresses that had sent or received a transaction~\cite{etherscan-charts}, not to mention continuous variables such as value, gas, and state changes~\cite{etherscan}.
This yields a combinatorial space of at least \(5 \times 10^{23}\).
Even if we assume each sender--receiver pair calls only one function, this space is too large to enumerate with brute-force techniques.

Organizing low-level artifacts into control-flow or dependency graphs is computationally expensive and often requires substantial effort to craft patterns and curate ground truth.
CLUE's execution property graph combines call hierarchy, asset flow, dynamic control flow, and data dependencies~\cite{qin2025enhancing} to capture various attack patterns.
However, using more on-chain features increases organization time, eroding their efficiency advantage.
It also requires substantial manual effort to craft patterns for each attack category from each on-chain feature type or combination.

Other detectors use more specific on-chain features.
TxSpector~\cite{zhang2020txspector} detects reentrancy attacks from call traces;
DeFiRanger~\cite{wu2023defiranger} and POMABuster~\cite{xi2024pomabuster} detect POMA from logs.
These techniques are limited to specific attack categories and rely on domain-specific patterns for attack detection.

\textbf{Challenge 2: Off-chain features.}
Off-chain features have limited availability and often require significant effort to clean, label, and verify for correctness and completeness.
Off-chain features include source code, historical transactions, and off-chain oracle data.
In our datasets, source code is available for 74.78\% of call traces, while function name coverage is 98.46\% (Section~\ref{sec:RQ1}). This 23.68-percentage-point gap makes source code a substantially less available input than names in attack detection.
Historical transactions have a cold-start issue: newly deployed protocols and emerging attack patterns have little or no labeled history, making learning-based detectors difficult to bootstrap.
Off-chain oracle data (\eg external price feeds) have availability and trust issues: oracle values may be missing or stale, and oracle manipulation or reliance on third-party providers can reduce robustness.

\textbf{Summary.}
We need a fine-grained yet compact view of runtime behavior that:
\begin{enumerate*}[label=(\roman*)]
    \item is efficiently extractable from on-chain artifacts (\eg call traces) with low overhead,
    \item captures high-level business semantics via protocol-grounded naming conventions (\eg interfaces), and
    \item minimizes reliance on off-chain artifacts and manual curation while supporting generalizable learning. 
\end{enumerate*}
\emph{Function names obtained by mapping selectors in call traces through the selector-to-name database provide a compact semantic view that is efficiently derived from runtime calls and grounded in high-level interface semantics, without heavyweight reconstruction or source code for the contracts appearing in each transaction.}

\subsection{Insights}
\label{sec:challenge-insight:insights}

Figure~\ref{fig:method-pipeline} previews how these insights inform \sysname{}'s sequence builder and classifier.

\textbf{Insight 1: Function names are derived from call traces but expose business semantics of applications.}
Developers of dApps choose descriptive function names that convey business logic. In the \cream{}, \texttt{supply}, \texttt{borrow}, and \texttt{liquidate} (Listing~\ref{listing:cream-finance-code}) represent the loan lifecycle (Figure~\ref{fig:cream-finance-illustration}). These names encode transaction semantics (Listing~\ref{lst:call-trace-tree}).

Even in attack transactions, adversaries cannot change deployed target contracts' function names because those names are immutable~\cite{buterin2013ethereum}.
Instead, attackers can only rename functions in the contracts they control. The former accounts for 98.70\% of call traces in attack transactions, while the latter accounts for 1.30\%. We study it empirically in Section~\ref{sec:RQ1} and its influence on attack evasion in Section~\ref{sec:RQ6}.

Although call traces record selectors rather than names, a database can map many selectors to their names.
While selector-to-signature is not one-to-one in theory, we can map a selector to the most likely name in practice (Section~\ref{sec:sequence-builder}), enabling robust semantic analysis.

\textbf{Insight 2: Function name sequences support learning-based methods to separate attacks from benign ones.}
Key differences between attack and benign transactions arise in function name sequence semantics. Modeling mapped function names as sequences yields strong features for detection without handcrafted rules or heavy analysis.

Note that our design focuses on semantic patterns from victim-side function names rather than relying on attacker-side function names.
In the \cream{}, atypical co-occurrences and ordering (\eg \texttt{borrow} followed by liquidation paths interleaved with attacker-created callbacks) contrast with a routine user flow (\texttt{borrow} \arrow~\texttt{repay} \arrow~\texttt{seize}).
These differences appear in victim-side function name sequences and are strong enough to support effective detection without source code or comments for the analyzed contracts, or additional execution properties. We evaluate this in Section~\ref{sec:RQ6}.

We hypothesize that such sequence-level deviations persist across blockchain application attack types because they reflect differences in business logic rather than low-level patterns.
We also hypothesize that learning-based methods can effectively capture these semantic differences in function name sequences to distinguish attacks from benign transactions without manual feature engineering or pattern crafting.
We experimentally validate these hypotheses in Section~\ref{sec:experiment}. %

%% file: sections/05-method.tex
\section{\sysname{} Design}
\label{sec:methodology}

In this section, we present \sysname{}, an on-chain application-level attack detector that uses raw call traces to derive function name sequences, without relying on contract source code or auxiliary transaction metadata during classification.
Figure~\ref{fig:method-pipeline} shows a high-level overview of \sysname{}'s architecture. 
\sysname{} comprises two components:

The \textbf{Sequence builder} reconstructs a call trace tree from the raw call traces of a transaction and maps function selectors to function name sequences.
It addresses \textbf{Challenge 1} by recovering high-level application semantics from low-level raw call traces through function name mapping and structured call trace reconstruction, as \textbf{Insight 1} suggests.

The \textbf{Classifier} encodes the function name sequence and applies a Transformer encoder followed by a linear head to predict whether each transaction is attack or benign.
It addresses \textbf{Challenge 2} by learning attack-related sequence patterns across the different attack categories, following \textbf{Insight 2}, without relying on handcrafted rules or state tracking.

\subsection{Sequence Builder}
\label{sec:sequence-builder}

The sequence builder reconstructs the hierarchical execution flow of a transaction from raw call traces and prepares an ordered function name sequence for learning.
It first curates a selector-to-name database, then formalizes transactions and links raw call traces into a call trace tree.
Finally, it maps 4-byte selectors to function names, yielding a representation that captures semantics and call-order context for classification.

\textbf{Build Selector-to-Name Database.}
We first extract the function signatures from the Sourcify smart contract dataset~\cite{sourcify-contract-api}.
For each extracted signature, we compute its 4-byte selector.
Because a function selector retains only the first four bytes of the signature's Keccak-256 hash, different function signatures may produce the same selector~\cite{solidity-abi-spec}.
To address this issue, for each selector, we rank candidate signatures by their prevalence across the Sourcify smart contract dataset~\cite{sourcify-contract-api}, and use canonical order as a tie-breaker.
Finally, we remove the parameter list from each function signature to obtain the corresponding function name, and eventually build the selector-to-name database.
All selectors without a corresponding function name in the database are mapped to a special \texttt{UNK\_SELECTOR} token, which is treated as a unique function name in the sequence.
Building the database is a one-time process rather than a per-transaction requirement.

\textbf{Define transaction and its call traces.}
In Section~\ref{sec:background:ethereum}, we introduced the concept of a transaction and its call traces. In this section, we formalize the definitions for precision.

We first define the entities and functions that can appear in transactions and call traces.
In a blockchain system, let 
\( \mathcal{U} \)
be a finite set of user addresses and
\( \mathcal{C} \)
be a finite set of smart contract addresses, where
\( |\mathcal{U}|=U \), \( |\mathcal{C}|=C \), and \( U, C \in \mathbb{N} \).
Let 
\( \mathcal{F}_{\text{pub}} \) and \( \mathcal{F}_{\text{ext}} \)
denote the public and external functions exposed through contract ABIs in the blockchain system. For each function
\( f \in \mathcal{F}_{\text{pub}} \cup \mathcal{F}_{\text{ext}} \),
we define 
\( f = (h_f, n_f, p_f) \),
where 
\( h_f \),
\( n_f \)
and 
\( p_f \)
denote the 4-byte function selector, the function name, and the parameter list of function
\( f \), respectively.
Denote
\( \mathcal{X} \)
as the set of transactions in the system. A transaction
\( x \in \mathcal{X} \)
is defined as a 4-tuple:
\[ x = ( s_x, r_x, v_x, t_x ), \]
where:
\begin{itemize}
    \item \( s_x \in \mathcal{U} \) is the sender of the transaction,
    \item \( r_x \in \mathcal{C} \) is the receiver contract, as we focus on transactions interacting with smart contracts,
    \item \( v_x \in \mathbb{N}_0 \) is the ETH value, and
    \item \( t_x \in \mathcal{T} \) is the root call trace generated when transaction \( x \) is executed by calling a public or external function.
\end{itemize}
We next define call traces, which are the nodes of the transaction call trace tree.
\( \mathcal{T} \)
represents the set of call traces in the system. A call trace 
\( t_k \in \mathcal{T} = \{ t_1, t_2, \ldots, t_T \} \)
(where 
\( k \in \{1, 2, \ldots, T\} \)
) is defined as the following tuple:
\begin{equation}
\label{eq:call_trace_definition}
t_k = ( f_k, s_k, r_k, v_k, \phi_k, \mathcal{C}_k )
\end{equation}
where:
\begin{itemize}
    \item \( f_k \in \mathcal{F}_{\text{pub}} \cup \mathcal{F}_{\text{ext}} \) is the public or external function invoked by call trace \( t_k \),
    \item \( s_k \in \mathcal{U} \cup \mathcal{C} \) is the caller of call trace \( t_k \),
    \item \( r_k \in \mathcal{C} \) is the callee contract of call trace \( t_k \),
    \item \( v_k \in \mathbb{N}_0 \) is the ETH value transferred in call trace \( t_k \),
    \item \( \phi_k \in \{\) \texttt{call}, \texttt{delegatecall}, \texttt{staticcall} \( \} \) is the call type of the call trace \( t_k \), and
    \item \( \mathcal{C}_k = (t_{k,1}, \ldots, t_{k,m_k}) \in \mathcal{T}^{m_k} \), with \( t_{k,j} \neq t_k \), denotes the ordered immediate child call traces associated with call trace \( t_k \).
\end{itemize}

The caller and callee fields are derived from raw trace metadata and are used to link traces into a tree; they are not consumed as classifier features in the default design.
For ease of presentation, we define functions 
\( S_{\text{tx}}: \mathcal{X} \to \mathcal{U} \)
and 
\( R_{\text{tx}}: \mathcal{X} \to \mathcal{C} \)
to obtain the sender (\eg dApp user) and the receiver contract of a transaction (\ie
\( S_{\text{tx}}(x) = s_x, \; R_{\text{tx}}(x) = r_x \)
for all 
\( x \in \mathcal{X} \)
).
Further, we define 
\( S_{\text{trc}}: \mathcal{T} \to \mathcal{U} \cup \mathcal{C} \)
and 
\( R_{\text{trc}}: \mathcal{T} \to \mathcal{C} \)
to obtain the caller and the callee contract of a call trace from raw trace metadata (\ie
\( S_{\text{trc}}(t_k) = s_k, \; R_{\text{trc}}(t_k) = r_k \)
for all
\( t_k \in \mathcal{T} \)
).

By organizing the call traces using this linked-node data structure, the complete call flow of a transaction naturally forms a call trace tree. In this structure, we define a function
\( O: \mathcal{X} \to \mathcal{T} \)
as the function that returns the root call trace of transaction \( x \), with \( O(x) = t_x \). The root call trace satisfies
\( S_{\text{trc}}(O(x)) = S_{\text{tx}}(x) \)
and
\( R_{\text{trc}}(O(x)) = R_{\text{tx}}(x) \).

Thus, we can now define the function 
\( D(\cdot\,;x): \mathcal{T} \to \mathbb{N}_0 \)
as the function that returns the depth of a call trace in transaction 
\( x \in \mathcal{X} \):
\begin{equation}
\label{eq:depth_definition} D(t;x) =
\begin{cases}
  0, & \text{if } t = O(x), \\
  D(t';x) + 1, & \text{if } \exists! \, t' \text{ in the trace tree of } x \\
                  & \text{such that } t \in \mathcal{C}_{t'}.
\end{cases}  
\end{equation}
From the recursive definition of the depth of a call trace in Eq.~(\ref{eq:depth_definition}), we can construct the call trace tree for each transaction \( x\in\mathcal{X} \) rooted at \( O(x) \).
Now, for a transaction \( x \in \mathcal{X} \), we further define the function 
\( I(\cdot\,;x): \mathcal{T} \to \mathbb{N} \)
as the function that returns the pre-order trace index in the call trace tree rooted at \( O(x) \).

\textbf{Map function selectors to names.}
After constructing the call trace tree, the sequence builder replaces each trace's selector \( h_f \) with its representative function name \( n_f \) using the selector-to-name database. After representative selection for colliding selectors, each selector in our database maps to one representative function name and parameter list.

Let \( \mathcal{N} \) be the finite vocabulary of function names, \( \mathcal{P} \) be the finite set of parameter lists, and \( \mathcal{H} \) be the finite set of observed function selectors. We define
\( M: \mathcal{H} \to \mathcal{N} \times \mathcal{P} \)
to map each observed 4-byte function selector to its representative function name and parameter list. For each call trace \( t \in \mathcal{T} \) with associated function \( f=(h_f,n_f,p_f) \), we compute
\[ M(h_f) = (n_f, p_f) \]
when \( h_f \in \mathcal{H} \). If \( h_f \notin \mathcal{H} \), we assign \( n_f=\text{\texttt{[UNK\_SELECTOR]}} \).

\subsection{Classifier}
\label{sec:classifier}

The classifier converts the call trace tree to a position-augmented function name sequence and applies a Transformer encoder with a linear head to predict attack or benign.

\textbf{Feature Engineering.}
\label{sec:feature-engineering}
Given a transaction \(x\), the sequence builder sorts calls by canonical \texttt{trace\_path}, yielding a deterministic pre-order call sequence \(\{t_i\}_{i=0}^{\mathcal{I}_x-1}\). Each function name is encoded by a frozen 8,192-token ByteLevel-BPE tokenizer and terminated by \texttt{[SEP]}.

\textbf{Model architecture.}
\label{sec:model}
The classifier uses a four-layer, eight-head pre-norm Transformer encoder~\cite{vaswani2017attention} with hidden width 256, SwiGLU feed-forward width 768, fused scaled-dot-product attention, and dropout 0.05. The final \texttt{[CLS]} representation passes through dropout and a linear two-class head.

\textbf{Training and inference.} 
\label{sec:training}
Given a labeled dataset \(\mathcal{D}=\{(\mathbf{s}_x,y)\}\), we train parameters \(\theta\) with cross-entropy and a balanced weighted sampler. At inference, checkpoint \(k\) predicts an attack when \(P_\theta(y=1\mid\mathbf{s}_x)\geq\tau_k\).

%% file: sections/06-experiment.tex
\section{Experiments}
\label{sec:experiment}

We evaluate \sysname{} by answering six research questions:
\begin{enumerate}[label=(RQ\arabic*)]
    \item How much coverage does a selector-to-name database provide for selectors in call traces?
    \item How does \sysname{} perform compared with existing approaches in detecting blockchain application attacks?
    \item Are \sysname{}'s detection results explainable?
    \item Can \sysname{} support real-time detection in popular blockchains?
    \item How much do function name sequences contribute to improved detection by \sysname{}?
    \item How robust is \sysname{} when directly attacker-controlled function names cannot be mapped?
\end{enumerate}
\subsection{Experiment Setup}
\label{sec:setup}

\textbf{Hardware and Software.}
Except for the post-fetch runtime benchmark, we run the experiments on a machine running Ubuntu 24.04 with an AMD Ryzen Threadripper 7960X and two NVIDIA RTX A5000 GPUs.
Call traces were obtained via QuickNode, and the selector-to-name database was constructed from the Sourcify smart contract dataset~\cite{sourcify-contract-api}. \sysname{} comprises 3,285 lines of Python code. Code, datasets, and scripts are detailed in the \hyperref[sec:appendix-open-science]{Open Science appendix}.

\textbf{Comparison.}
We select two state-of-the-art detectors and one call trace length heuristic as primary baselines for comparison with \sysname{}:
\begin{itemize}
\item Hunting~\cite{wu2025hunting} is an off-chain graph-neural-network-based transaction-level detector that learns transaction-graph representations through transaction reconstruction and uses source-code function comments as pretraining supervision.
\item CLUE~\cite{qin2025enhancing} detects reentrancy and POMA transactions from on-chain data by crafting patterns over an execution property graph that combines call traces, control flow, and token flow. 
\item Length heuristic is a simple baseline that classifies transactions as attacks if their call trace length exceeds a threshold. We set the threshold to 22, which yields the best FNR in our datasets under a FPR budget of 1\%.
\end{itemize}

We choose Hunting and CLUE because both operate at the application level, do not require smart contract source code at detection time, and provide runnable artifacts with transaction-level outputs, enabling comparison with \sysname{} on shared datasets. 
We use Hunting's pretrained model weights provided by its authors.
Hunting compares with DeFiScanner~\cite{wang2022defiscanner}, DEFIER~\cite{su2021evil}, and MoTS~\cite{wu2023know}, and reports higher precision, recall, and AUC than those detectors.
Likewise, CLUE compares with Sereum~\cite{rodler2018sereum} and TxSpector~\cite{zhang2020txspector} for reentrancy, reporting a lower FPR than both and a lower FNR than TxSpector; for price manipulation, it reports a lower FNR than DeFiRanger~\cite{wu2023defiranger}, FlashSyn~\cite{chen2024flashsyn}, DeFiTainter~\cite{kong2023defitainter}, and DeFort~\cite{xie2024defort}.

\input{assets/table-dataset.tex}

\textbf{Datasets.}
We construct our dataset from two sources, \ie DeFiHackLab~\cite{DeFiHackLabs} and Ethereum, and use it for training and evaluation. All transactions in our dataset have an attack or benign label. Table~\ref{tab:dataset} gives the dataset breakdown.

At the March 27, 2026 snapshot used in this work, DeFiHackLab~\cite{DeFiHackLabs} was a community-driven open-source project listing 688 incidents over 8 years, ranging from July 19, 2017 to March 27, 2026.
Each incident comes with the attack transaction hash, attacker and victim address, incident report URL, and the category based on the type of attack. 
We exclude incidents without metadata, which leaves 480 incidents.
We further exclude incidents outside our scope (Section~\ref{sec:threat-model}), \ie bridge attacks, governance attacks, incorrect validation, misconfiguration, overflow, precision loss, private-key compromise, signature verification, social engineering, storage collision, weak randomness, and other unspecified attacks, which leaves 424 in-scope incidents.
We collect all attack transactions from these 424 incidents, yielding 14,611 transactions with attack labels.
This attack set represents the attacks against the victims.
Similarly, we collect all victim addresses and their historical transactions, and then remove any transactions that are also in the above attack set, yielding 134,665 transactions with benign labels. 
This victim set represents the normal transactions of the victims.
Each incident has at least one attack transaction, and multiple victim transactions.
For training and evaluation, we split by incident, keeping each incident's transactions together to prevent leakage.

We collect all transactions from Ethereum mainnet from January 1, 2024 through December 31, 2025.
Similar to the victim set, we remove any transactions that are also in the above attack set, yielding 537,069,442 transactions.
This benign set represents normal blockchain-user transactions.

\textbf{Model training details.}
We partition all 424 incidents in the DeFiHackLab dataset into five category-stratified folds, with their attack sets (as attack) and victim sets (as benign). We also randomly sample a subset of 2048 transactions from the Ethereum dataset for each fold (as benign). We train five checkpoints. Each checkpoint fits four folds and uses the fifth for evaluation. Consequently, no checkpoint fits all 424 incidents in the train set simultaneously, avoiding incident leakage between its fit and selection folds. We report the five checkpoints' average metrics on their evaluation folds.

\subsection{RQ1: Coverage of Function Name}
\label{sec:RQ1}

We evaluate the coverage of function names Vs. source code features on selectors in the call traces.
Function name coverage matters because \sysname{} relies on function names to represent transaction behavior.
If function name coverage is low, many calls are represented as \texttt{[UNK\_SELECTOR]}, removing information about which functions were invoked.
Source code coverage matters because detectors that require source code cannot extract their required features when the called contract's source code is unavailable.
Low coverage of either feature can therefore degrade the detection accuracy of detectors that rely on that feature.

The transactions evaluated are from the DeFiHackLab dataset, which contains 14,611 attack transactions and 134,665 victim-history transactions (totalling 149,276 transactions). 
We conduct the experiment as follows:
\begin{enumerate*}[label=(\arabic*)]
    \item We calculate call traces in the 149,276 transactions, excluding calls without selectors (\eg EVM builtin functions), yielding 5,632,984 call traces.
    \item For each call trace, we look up the function name in our selector-to-name database; if the function name is found, we count the call trace as having a function name.
    \item We check whether the called contract has source code available in the Sourcify dataset~\cite{sourcify-contract-api}; if so, we count the call trace as having source code.
    \item We calculate both coverages as percentages over the same 5,632,984 call traces.
\end{enumerate*}

\textbf{Function names have 23.68 percentage points higher availability than source code.} 
The selector-to-name database provides function names for 5,546,485 call traces (98.46\%), whereas source code is available for 4,212,186 call traces (74.78\%), a difference of 23.68 percentage points.
Moreover, function names are available for 1,352,905 of the 1,420,798 call traces (95.22\%) whose source code is unavailable.
Thus, \sysname{} retains function-name information for nearly all call traces, while source-dependent detectors lack their required features for 25.22\% of calls.

\textbf{98.70\% of call traces in attack transactions invoke victim functions.} We assume that functions in contracts deployed and called by attackers are attack functions whose names are controlled by the attacker. Across 3,451,926 call traces from our attack dataset, only 44,716 calls (1.30\%) invoke attacker-controlled functions; the remaining 3,407,210 calls (98.70\%) invoke victim contracts. Section~\ref{sec:RQ6} evaluates how attacker evasion affects detection accuracy.

\insight{Coverage of Function Name.}{Our selector-to-name database provides function names for 98.46\% of calls in the DeFiHackLab dataset, which is 23.68 percentage points higher than source-code coverage. This implies that function names are able to provide much higher coverage for a transaction than source code, and can hence support detection of a much larger proportion of attacks.}

\subsection{RQ2: Detection Accuracy}
\label{sec:RQ2}

\input{assets/table-rq1-fnr-overall.tex}

\textbf{False negative rate (FNR).}
FNR is the ratio of false negative incidents to all 424 labeled incidents in the attack dataset.
To prevent incident leakage, we place every incident's attack and victim-history transactions in a single fold and evaluate that incident only with the checkpoint fitted on the other four folds.
We report the average FNR across the five folds.
Table~\ref{tab:rq2-fnr-overall} presents this incident-level result for \sysname{}, the Length Heuristic, CLUE, and Hunting.
We do not use the victim or benign dataset for evaluating FNR, as it does not have known attack transactions (recall that we removed all transactions that matched the attack dataset).
\emph{Overall, \sysname{} is highly effective at detecting blockchain application attacks with an FNR of 1.56\% on the attack dataset.
In comparison, the length heuristic, CLUE, and Hunting obtain FNR values of 31.31\%, 70.77\%, and 83.17\%, respectively, on the same set.
All three baselines thus have higher FNR than \sysname{}. }

\emph{Why does Hunting report high FNR?} 
One plausible explanation is selection and temporal bias in Hunting's source code and comment-supervised pretraining.
Hunting's transaction-comment contrast (TCC) feature captures the similarity between transactions and source code/comments.
It requires the transaction to be similar to the matched code/comments and vice versa, and thus may flag an attack because it deviates from the code/comments, when such artifacts are available.   
Removing TCC decreases its reported macro F1 score by 61.4\%~\cite{wu2025hunting}.
Furthermore, source code is available for 74.78\% calls according to our empirical study in Section~\ref{sec:RQ1}.
Pretraining only from transactions with source code and comments may lack source-unavailable behavior, while \sysname{} retains function name features for nearly all such behaviors.
This selection effect may be compounded by temporal shift. Hunting's pretraining corpus samples Ethereum blocks 1--18,000,000, which only includes contracts deployed before May 2023, whereas our attack dataset includes attacks through March 2026.
Both biases can contribute to Hunting's false negatives.
Consistent with these hypotheses, Appendix~\ref{sec:appendix-baseline-comparison} reports that Hunting's FNR is lower on its own dataset.

\emph{Why does CLUE report high FNR?} 
Similar issues may also affect CLUE, as its EPG patterns are derived from and evaluated on a dataset from Zhou \etal{}~\cite{zhou2023sok} that contains blockchain application attacks up to 2022. Changes in smart contract development practices and the emergence of new attack vectors may render the EPG patterns less effective over time, leading to higher FNR. These results suggest that detectors relying on fixed historical patterns or source-code-derived features can lose effectiveness as smart contract practices and attack vectors evolve. In contrast, \sysname{} uses function name sequences derived from standard interface conventions rather than evolving contract implementations (Section~\ref{sec:discussion}).

\emph{Why does the length heuristic have a lower FNR than Hunting and CLUE?}
The length heuristic achieves an FNR of 31.31\%, lower than CLUE (70.77\%) and Hunting (83.17\%), indicating that long call traces capture some transaction-complexity signal associated with attacks.
However, call trace length is not attack-specific: benign transactions can also produce long traces, while attacks with shorter traces are missed.
At the selected threshold of 22, the heuristic still misses nearly one-third of attack incidents and produces an FPR of 0.93\%, approximately 550 times \sysname{}'s 0.0017\% FPR; in comparison, \sysname{} achieves an FNR of only 1.56\% (Table~\ref{tab:rq2-fnr-overall}).
Thus, though the length heuristic outperforms Hunting and CLUE in FNR, its substantially worse FNR--FPR tradeoff shows that length is not an accurate discriminator of attacks.

\emph{Comparison with baselines on their own datasets.}
To reduce the effect of temporal and dataset shift on the comparison, we conduct a cross-dataset evaluation using the CLUE dataset~\cite{qin2025enhancing} and the DAppFL dataset used by Hunting~\cite{wu2024dappfl}. 
We run CLUE, Hunting, and the frozen \sysname{} checkpoints from our primary evaluation on both CLUE and DAppFL datasets and measure their FNR.
We reuse the \sysname{} checkpoints without retraining.
See Appendix~\ref{sec:appendix-baseline-comparison} for results.

\insight{FNR.}{\sysname{} achieves an FNR of 1.56\% on the attack dataset, lower than length heuristics (31.31\%), Hunting (83.17\%) and CLUE (70.77\%).}

\textbf{False positive rate (FPR).}
FPR is the ratio of false positives to transactions treated as benign.
We run \sysname{} on the benign dataset.
We exclude the attack dataset from the FPR evaluation because it contains only attack transactions.

\emph{Out of 537,069,442 transactions in the benign dataset, \sysname{} flags 9,228 transactions as potential attacks, reporting an FPR of 0.0017\%.}
Across the benign dataset's 24-month observation period, these 9,228 positives translate to 12.6 alerts per day, which is a tractable triage load for operators, and an operational strength of \sysname{}.
We note that this FPR is a conservative upper bound under our validation methodology, as the benign dataset may contain undisclosed attacks not catalogued in public databases.

Figure~\ref{fig:rq4-fp-categories} summarizes the function-name categories assigned to the false-positives predictions. 
We find that one of these 9,228 transactions is a real attack by web searching, and is hence not a false-positive.
For the remaining 9,227 transactions, Unknown selector accounts for 4,516 transactions (48.94\%) and Token accounts for 3,300 (35.76\%); together, they account for 84.71\%.
The remaining categories are Execution (5.65\%), Other or custom functions (4.61\%), Flashloan and leverage (2.57\%), Swap and liquidity (1.14\%), Callback (0.70\%), and Lending (0.63\%).
We discuss false-positives analysis further in Appendix~\ref{sec:appendix-false-positives}.

\input{assets/fig-rq4-fp-categories.tex}

CLUE reports FPRs between 0.52\% and 1.49\% on its own benign datasets~\cite{qin2025enhancing}.
Hunting does not report its FPR~\cite{wu2025hunting}, precluding comparison with \sysname.

\insight{FPR.}{\sysname{}'s FPR is 0.0017\%.}

\textbf{Runtime overhead.}
Table~\ref{tab:rq2-fnr-overall} also reports the average detection runtime of \sysname{}, Hunting, and CLUE.
The average end-to-end runtime for \sysname{} from trace availability to classification is 24.90~\emph{milliseconds} per transaction, with a maximum of 1{,}819.032 milliseconds and a P95 of 50.03 milliseconds.
On the same dataset, Hunting and CLUE average 59.53 and 275.43 seconds, respectively.

\emph{Why do Hunting and CLUE have high time overhead?} 
The runtime differences among \sysname{}, CLUE, and Hunting appear to come from sensitivity to transaction complexity and differences in feature-extraction pipelines and implementation parallelism.
\sysname{}, Hunting, and CLUE exhibit different scaling with transaction complexity. Both \sysname{} and Hunting employ a neural network as the detector backbone; self-attention has quadratic complexity in the input sequence length~\cite{dolga2024latent}, although the observed end-to-end runtime also depends on preprocessing, batching, hardware, and implementation. In contrast, CLUE constructs an execution property graph (EPG) for each transaction and mines subgraph patterns, which can be more sensitive to the structural complexity of the EPG rather than only the length of the call trace sequence. This helps explain its higher overhead in our measurements.

\insight{Time Overhead.}{\sysname{} has an average detection time of 24.90 ms and a P95 of 50.03 ms.}

\subsection{RQ3: Explainability of Detection}
\label{sec:RQ3}

Explainability of detection matters for \sysname{} because analysts need to understand which function names support its attack predictions.
We explain each positive prediction by comparing \emph{the attack mechanism in the incident report} with the function names that contribute positively to the prediction.
If these function names align with the exploit mechanism, then the prediction is supported by behavior consistent with the report, providing analysts with an intuitive explanation.
To identify the positively contributing function names, we apply post-hoc Layer Integrated Gradients (LIG)~\cite{sundararajan2017axiomatic,kokhlikyan2020captum}.

We start with the 424 incidents from our attack dataset.
We search incident reports from official or reputable sources, yielding 184 reports for analysis - the other 240 incidents lack a qualifying report.
For each of the 184 incidents with reports, we pick one attack transaction each from our attack dataset.
Note that these 184 attack transactions are predicted as positive (attack) by \sysname{} and hence no false negative incidents are included in this explainability analysis.
We analyze the 184 report and transaction pairs as follows:
\begin{enumerate*}[label=(\arabic*)]
    \item For the transaction in the pair, we aggregate LIG by function name and retain every unique function name with positive attribution.
    \item For the report, we use the report's exploit mechanism evidence, mask other details, and encode the report evidence and positive function names with a pretrained sentence Transformers encoder~\cite{sentence-transformers-all-mpnet-base-v2}, which is not fine-tuned for this specific task.
    \item We define the alignment score as the percentile rank of the true report among all 184 reports for the same attribution. A higher score means that \sysname{}'s positive predictions fit their report more specifically.
\end{enumerate*}

\input{assets/fig-rq3-tp-alignment-scatter.tex}

Figure~\ref{fig:rq3-tp-alignment-scatter} shows the distribution of alignment scores.
The median alignment score is 74.86\%.

\input{assets/fig-rq3-cream-explainability.tex}

\textit{Case Study.}
Figure~\ref{fig:rq3-cream-explainability} illustrates this agreement for the \cream{}, which has an alignment score of 97.27\%.
The top row shows three call-trace subtrees; red nodes denote positively attributed calls, and the adjacent labels identify their function names.
The bottom row presents corresponding excerpts from the official incident report~\cite{cream-finance-blog}.
Highlighted report phrases are connected to the relevant calls above, making the agreement between \sysname{}'s attribution and the documented exploit mechanism explicit.
First, the initial-borrow subtree contains two positively attributed \texttt{borrow} calls, matching the report's identification of the protocol's \texttt{borrow} function as vulnerable.
Second, the reentrant-borrow subtree contains the positively attributed path \texttt{tokensReceived}$\rightarrow$\texttt{borrow}, matching the report's description of AMP's \texttt{tokenReceived} hook triggering a nested second \texttt{borrow} before the initial borrow completes.
Third, the liquidation subtree's positively attributed \texttt{liquidateBorrow}, \texttt{repayBorrowAllowed}, and \texttt{seize} calls align, respectively, with the report's account of the attacking contract liquidating its counterpart, repaying AMP, and seizing its collateral~\cite{cream-finance-blog}.

\insight{Explainability.}{Across 184 reports, \sysname{}'s predictions have 74.86\% median alignment.}

\subsection{RQ4: Real-Time Detection}
\label{sec:RQ4}

\input{assets/table-real-world-deployment-blockchain-throughput.tex}

For real-time deployment, detection must finish before transaction commitment (\ie block production). Otherwise, the deployed method will not be able to detect potential attacks in real time.
We randomly pick a month's worth of data (\ie September 2025) from DeFiLlama for analysis.
DeFiLlama is a well-known DeFi analytics platform that collects incidents and losses across multiple blockchains.

Table~\ref{tab:real-world-deployment-blockchain-throughput} summarizes the block times and transaction throughputs of the seven most vulnerable blockchains~\cite{defillama-hack}.
\sysname{}'s warmed post-fetch latency has a mean of $24.90$ milliseconds and a P95 of $50.03$ milliseconds, both of which are below the shortest block time in Table~\ref{tab:real-world-deployment-blockchain-throughput} (Arbitrum, 0.25 seconds). However, the measured maximum is 1{,}819.032 milliseconds, and the benchmark does not measure trace acquisition, full-block throughput, or production traffic.
\emph{Thus, typical per-transaction latency is compatible with block production.}

\sysname{} is designed for deployment by block builders or validators who execute candidate transactions locally before block construction.
The measured $24.90$~ms mean spans the time from trace availability to serialized classification output, and excludes network latency and block confirmation.
This trace-to-output interval is the standard way to estimate detection overhead at the validator side, where traces are available from local transaction execution.

\insight{Real-Time Detection}{\sysname{}'s latency supports real-time detection on popular blockchains.}

\subsection{RQ5: Ablation Study}
\label{sec:RQ5}

We study how \sysname{}'s two input signals contribute to detection and how an older-to-recent split affects the result. 
The feature experiment compares the full model with two retrained ablations: function names only and tree position only.
The function-names-only arm retains the frozen function name tokens and masks tree and order inputs. 
The tree-position-only arm replaces each retained name with a constant call marker while retaining position encoding.

To compare feature arms at the operating point, the ablation study uses the same FPR target. At each epoch, selection minimizes false negatives under that target.

\input{assets/fig-rq5-feature-ablation.tex}

Figure~\ref{fig:rq5-feature-ablation} compares the incident FNR of the full model with the two feature ablations and the older-to-recent temporal ablation.
Recall that in RQ2, \sysname{} reports an FNR of 1.56\% for all incidents.
Removing tree position raises the FNR to 3.30\%, while removing function names raises it to 8.73\%. 
Thus, both signals contribute, and function names provide the stronger standalone signal under these ablations.
We evaluate call trace length as the length-heuristic baseline in RQ2 (Table~\ref{tab:rq2-fnr-overall}); it yields a 31.31\% incident FNR at its threshold, worse than both feature ablations.

We fit the chronological diagnostic on 337 older incidents and evaluate 87 recent ones.
It misses 1 of 87 (1.15\% FNR).

\insight{Ablation Study.}{Function names are the key features for detection: removing them raises the FNR to 8.73\%. Function names are not sensitive to temporal changes: its ablation has a 1.15\% FNR.}

\subsection{RQ6: Evasion}
\label{sec:RQ6}

Since \sysname{} relies on function names and call trace-tree structure, we examine its robustness when names in the directly controlled attacker contract are entirely unavailable. We then use the \cream{} to illustrate why removing victim-side calls requires exploit-specific reasoning.

\textbf{Making direct-attacker names unknown.} 
Attackers can rename functions in contracts they control, but they cannot rename functions in already-deployed victim contracts. 
As reported in RQ1, calls to the former account for only 1.30\% of all call traces across our attack dataset.
We replace every attacker-controlled function name with the canonical unknown-selector token \texttt{[UNK\_SELECTOR]} and rerun inference \emph{without retraining}.
This removes lexical information for those names, modeling a setting in which they are absent from the frozen selector database.

This perturbation raises \sysname{}'s FNR from 1.56\% to 2.26\%, a 0.70 percentage-point increase.
Thus, complete loss of lexical information for these names has only limited adverse effect on sensitivity.
The result is consistent with the RQ1 call-share measurement: 
98.70\% of retained calls invoke non-attacker contracts under the attribution rule.
These calls retain function name evidence even when every directly attacker-controlled name becomes unknown.

\textbf{Removing call traces.} 
Whether a call can be removed while preserving an exploit requires case-specific analysis of the attack's execution and profitability. 
We use the minimized \cream{} trace only to illustrate trace-removal constraints, not as a quantitative experiment.

\input{assets/list-call-trace-tree-min.tex}

Listing~\ref{lst:call-trace-tree-min} retains the victim-side operations marked [C]. In particular, the attacker must retain the two victim-side borrows as well as the liquidation, seizure, and redemption operations that realize the exploit and its profit. The attacker must retain \texttt{AMPMarket.liquidate}: without liquidation, the supplied collateral remains exposed to other liquidators and the exploit is not profitable.

By contrast, preparatory calls such as \texttt{Uniswap.swap}, \texttt{WETH.withdraw}, and \texttt{ETHMarket.supply} can be performed separately when obtaining collateral. 
Calls from \texttt{WETH.deposit} onward can likewise be omitted because they transfer realized profit to the attacker's wallet. 
The remaining trace illustrates why a successful exploit may still expose recognizable victim-side business semantics even after removable preparation and profit-transfer calls are excluded.

\insight{Evasion.}{Replacing attacker-controlled function names raises FNR from 1.56\% to 2.26\%. Removing victim-side calls requires exploit-specific reasoning to preserve exploit execution and profitability.}

%% file: assets/table-dataset.tex
\begin{table}[!t]
    \caption{Breakdown of the labeled transaction datasets.}
    \label{tab:dataset}
    \centering
    \footnotesize
    \setlength{\tabcolsep}{3pt}
    \begin{tabular}{@{}lllrr@{}}
        \toprule
        \textbf{Name} & \textbf{Source} & \textbf{Label} & \textbf{\#Txns} & \textbf{Duration} \\
        \midrule
        Attack & DeFiHackLab & Attack & 14,611 & 8 yrs \\
        Victim & DeFiHackLab & Benign & 134,665 & 8 yrs \\
        Benign & Ethereum & Benign & 537,069,442 & 2 yrs \\
        \bottomrule
    \end{tabular}
\end{table}

%% file: assets/table-rq1-fnr-overall.tex
\begin{table}[!t]
    \centering
    \caption{FNR (\%), FPR (\%), and average detection time.}
    \label{tab:rq2-fnr-overall}
    \footnotesize
    \begin{tabular}{lrrr}
        \toprule
        \textbf{Method} & \textbf{FNR} & \textbf{FPR} & \textbf{Mean time} \\
        \midrule
        \sysname{} & 1.56 & 0.0017 & \underline{24.90 ms} \\
        Length Heuristics & 31.31 & 0.93 & n/a \\
        Hunting~\cite{wu2025hunting} & 83.17 & n/a & 59.53 s \\
        CLUE~\cite{qin2025enhancing} & 70.77 & n/a & 275.43 s \\
        \bottomrule
    \end{tabular}
\end{table}

%% file: assets/fig-rq4-fp-categories.tex
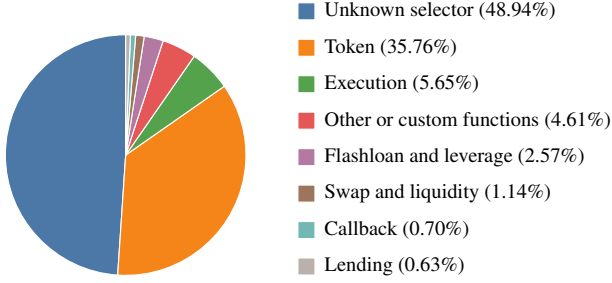
\begin{figure}[!t]
    \centering
    \definecolor{rqfpblue}{HTML}{4C78A8}
    \definecolor{rqfporange}{HTML}{F58518}
    \definecolor{rqfpgreen}{HTML}{54A24B}
    \definecolor{rqfpred}{HTML}{E45756}
    \definecolor{rqfppurple}{HTML}{B279A2}
    \definecolor{rqfpbrown}{HTML}{9D755D}
    \definecolor{rqfpteal}{HTML}{72B7B2}
    \definecolor{rqfpgray}{HTML}{BAB0AC}
    \tikzset{rqfpslice/.style={draw=white,line width=0.45pt}}
    \resizebox{0.96\columnwidth}{!}{%
    \begin{tikzpicture}[font=\footnotesize]
        \pie[
            sum=100,
            radius=1.65,
            rotate=90,
            text=legend,
            hide number,
            style={rqfpslice},
            color={rqfpblue,rqfporange,rqfpgreen,rqfpred,
                   rqfppurple,rqfpbrown,rqfpteal,rqfpgray}
        ]{
            48.94/{Unknown selector (48.94\%)},
            35.76/{Token (35.76\%)},
            5.65/{Execution (5.65\%)},
            4.61/{Other or custom functions (4.61\%)},
            2.57/{Flashloan and leverage (2.57\%)},
            1.14/{Swap and liquidity (1.14\%)},
            0.70/{Callback (0.70\%)},
            0.63/{Lending (0.63\%)}
        }
    \end{tikzpicture}%
    }
    \caption{Top positively attributing function name categories for the 9,227 FP transactions.}
    \label{fig:rq4-fp-categories}
\end{figure}

%% file: assets/fig-rq3-tp-alignment-scatter.tex
\newif\ifrqtpLegacyAlignmentFigure
\rqtpLegacyAlignmentFigurefalse

\ifrqtpLegacyAlignmentFigure
\begin{figure}[!t]
    \centering
    \definecolor{rqtpalignment}{HTML}{0877B5}
    \definecolor{rqtpgrid}{HTML}{D7D7D7}
    \resizebox{\columnwidth}{!}{%
    \begin{tikzpicture}[x=0.16cm,y=0.032cm]
        \foreach \y in {0,25,50,75,100} {
            \draw[rqtpgrid,line width=0.35pt] (0,\y) -- (36,\y);
            \node[anchor=east,font=\footnotesize] at (-0.8,\y) {\y\%};
        }
        \foreach \x in {0,5,10,15,20,25,30,35} {
            \draw[rqtpgrid,line width=0.35pt] (\x,0) -- (\x,100);
            \node[anchor=north,font=\footnotesize] at (\x,-3.5) {\x};
        }

        \foreach \rqtx/\rqty in {
            1/100,5/96.17,9/92.9,11/99.45,4/93.99,4/98.91,
            25/87.98,8/61.2,11/94.54,20/53.01,17/63.39,14/70.49,
            19/22.4,12/95.63,15/100,10/31.69,7/19.67,31/97.27,
            10.82/99.45,12/93.99,11/39.34,5/87.98,0.82/100,9/38.8,
            14/79.78,23/100,14/88.52,11/22.4,5/26.78,10/54.1,
            11/53.01,9/43.17,1/86.89,16/86.34,9/89.62,5/92.9,
            19/27.32,15/98.91,6/67.76,9/93.99,17/67.21,3/38.25,
            7/0,11/100,8/91.8,19/10.38,12/78.14,9/69.95,
            11/74.86,13/16.39,8/100,11/66.12,8/98.91,8/80.87,
            8/12.57,10/18.03,9/90.16,28/8.74,8/76.5,18/80.33,
            13/97.27,12/71.04,7.82/100,20/18.03,11/5.46,12/37.16,
            29/66.67,3/100,22/85.25,11/4.37,14/23.5,19/25.68,
            12/1.09,8/80.33,29/100,13/59.56,25/73.77,15/37.16,
            4/95.63,6/55.19,7/93.99,6/68.31,25/62.84,8/71.58,
            5/81.42,13/72.68,22/48.63,12/31.15,18/82.51,13.82/79.78,
            15/54.1,10/93.44,20/49.18,9/58.47,22/38.25,11/91.26,
            7/62.84,2/83.06,13/96.17,21/57.38,14/30.05,8/70.49,
            12/85.25,15/95.63,35/94.54,18/99.45,8/94.54,15/3.28,
            10/28.42,9/74.86,20/88.52,1/31.69,9/95.63,2/46.99,
            13/87.43,12/60.11,14/45.36,20/79.23,3/0,17/84.15,
            9/24.04,3/49.18,10/50.82,9/27.87,8/27.87,8/10.93,
            22.82/100,21/90.71,4/99.45,4/87.98,18/91.26,7/61.75,
            21/47.54,10/24.04,5/16.39,13/91.8,6/12.02,7/89.07,
            9/56.28,3/27.32,1/83.06,16/79.78,12/81.42,12/87.98,
            8/84.7,16/99.45,5/36.61,14.82/100,14/69.4,4/96.72,
            8/85.79,12/55.19,14/64.48,16/89.62,5/92.35,10/47.54,
            18/85.25,1/25.68,8.18/100,2/98.36,5/85.25,14/3.83,
            14/73.22,10/100,23/7.65,6/83.06,8.82/93.99,9/20.77,
            2/74.86,24/75.96,12/20.77,13/100,10/92.9,6/100,
            18/87.98,13/87.98,14/97.81,13/51.37,5/71.58,16/61.2,
            20/95.08,12/54.1,19/63.39,10/73.22
        } {
            \fill[rqtpalignment,opacity=0.68] (\rqtx,\rqty) circle[radius=0.85pt];
        }

        \draw[black!65,densely dashed,line width=0.65pt] (0,74.86) -- (36,74.86);
        \node[anchor=east,font=\footnotesize,fill=white,inner sep=1.2pt] at (35.5,78.5) {Median alignment score: 74.86\%};
        \draw[black!65,densely dashed,line width=0.65pt] (11,0) -- (11,100);
        \node[anchor=west,font=\footnotesize,fill=white,inner sep=1.2pt] at (11.5,5) {Median positive names: 11};
        \draw[->,line width=0.55pt] (0,0) -- (37.5,0);
        \draw[->,line width=0.55pt] (0,0) -- (0,105);
        \node[anchor=north,font=\footnotesize] at (18,-13) {Eligible positive function names (count)};
        \node[rotate=90,anchor=south,font=\footnotesize] at (-5.7,50) {Alignment score};
    \end{tikzpicture}%
    }
    \caption{Contrastive semantic alignment for 184 assessable incident representatives.}
    \label{fig:rq3-tp-alignment-scatter}
\end{figure}
\else
\begin{figure}[!t]
    \centering
    \definecolor{rqtpalignment}{HTML}{0877B5}
    \definecolor{rqtpgrid}{HTML}{D7D7D7}
    \resizebox{\columnwidth}{!}{%
    \begin{tikzpicture}[x=0.075cm,y=0.065cm]
        \foreach \y in {0,10,20,30,40,50} {
            \draw[rqtpgrid,line width=0.35pt] (0,\y) -- (100,\y);
            \node[anchor=east,font=\footnotesize] at (-2,\y) {\y};
        }
        \foreach \x in {0,20,40,60,80,100} {
            \draw[rqtpgrid,line width=0.35pt] (\x,0) -- (\x,55);
            \node[anchor=north,font=\footnotesize] at (\x,-2.2) {\x\%};
        }

        \foreach \xlo/\xhi/\rqtn in {
            0/10/9,10/20/9,20/30/15,30/40/11,40/50/8,
            50/60/13,60/70/16,70/80/19,80/90/30,90/100/54
        } {
            \filldraw[fill=rqtpalignment,fill opacity=0.78,draw=white,line width=0.55pt]
                (\xlo,0) rectangle (\xhi,\rqtn);
        }

        \draw[black!70,densely dashed,line width=0.8pt] (74.86,0) -- (74.86,55);
        \node[anchor=east,font=\footnotesize,fill=white,inner sep=1.2pt]
            at (73.2,51.5) {Median: 74.86\%};

        \draw[->,line width=0.55pt] (0,0) -- (104,0);
        \draw[->,line width=0.55pt] (0,0) -- (0,58);
        \node[anchor=north,font=\footnotesize] at (50,-9.5) {Alignment score};
        \node[rotate=90,anchor=south,font=\footnotesize] at (-10,27.5) {Incident count};
    \end{tikzpicture}%
    }
    \caption{Distribution of alignment scores for 184 assessable incident pairs.}
    \label{fig:rq3-tp-alignment-scatter}
\end{figure}
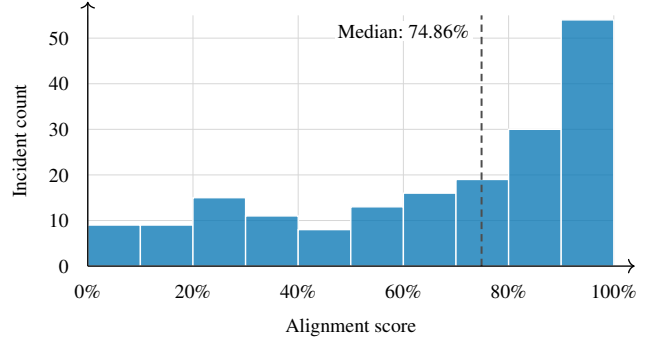
\fi

%% file: assets/fig-rq3-cream-explainability.tex
\begin{figure*}[!t]
    \centering
    \begin{minipage}{\textwidth}
    \centering
    \definecolor{rqcreamgold}{HTML}{C98200}
    \definecolor{rqcreamborder}{HTML}{AEB4B8}
    \definecolor{rqcreamdivider}{HTML}{D8DBDD}
    \definecolor{rqcreampeach}{HTML}{FFD3C2}
    \definecolor{rqcreamheat00}{HTML}{053061}
    \definecolor{rqcreamheat01}{HTML}{2166AC}
    \definecolor{rqcreamheat02}{HTML}{4393C3}
    \definecolor{rqcreamheat03}{HTML}{92C5DE}
    \definecolor{rqcreamheat04}{HTML}{D1E5F0}
    \definecolor{rqcreamheat05}{HTML}{F7F7F7}
    \definecolor{rqcreamheat06}{HTML}{FDDBC7}
    \definecolor{rqcreamheat07}{HTML}{F4A582}
    \definecolor{rqcreamheat08}{HTML}{D6604D}
    \definecolor{rqcreamheat09}{HTML}{B2182B}
    \definecolor{rqcreamheat10}{HTML}{67001F}
    \definecolor{rqcreamnodeborder}{HTML}{666D73}
    \pgfdeclarehorizontalshading{rqcreamheatmap}{1cm}{%
        color(0bp)=(rqcreamheat00);%
        color(25bp)=(rqcreamheat00);%
        color(30bp)=(rqcreamheat01);%
        color(35bp)=(rqcreamheat02);%
        color(40bp)=(rqcreamheat03);%
        color(45bp)=(rqcreamheat04);%
        color(50bp)=(rqcreamheat05);%
        color(55bp)=(rqcreamheat06);%
        color(60bp)=(rqcreamheat07);%
        color(65bp)=(rqcreamheat08);%
        color(70bp)=(rqcreamheat09);%
        color(75bp)=(rqcreamheat10);%
        color(100bp)=(rqcreamheat10)%
    }
    \resizebox{\textwidth}{!}{%
    \begin{tikzpicture}[x=1cm,y=1cm]
        \tikzset{
            rqtitle/.style={anchor=west,inner sep=0pt,font=\rmfamily\footnotesize},
            rqcopy/.style={anchor=base west,inner sep=0pt,font=\rmfamily\footnotesize},
            rqlegendlabel/.style={inner sep=0pt,font=\rmfamily\fontsize{5}{5}\selectfont},
            rqlegendtick/.style={anchor=north,inner sep=0pt,font=\rmfamily\fontsize{4}{4}\selectfont},
            rqpeach/.style={fill=rqcreampeach,rounded corners=0.035cm,inner xsep=0.045cm,inner ysep=0.025cm},
            rqmap/.style={draw=rqcreamgold,line width=0.42pt,line cap=round,line join=round},
        }

        \begin{scope}[yshift=1.20cm]
        \foreach \x in {0,6.05,12.10} {
            \fill[white] (\x,1.08) rectangle +(5.75,3.47);
            \draw[rqcreamborder,line width=0.45pt,rounded corners=0.05cm]
                (\x,1.08) rectangle +(5.75,3.47);
            \draw[rqcreamdivider,line width=0.40pt]
                (\x+0.18,3.48) -- (\x+5.57,3.48);
        }
        \end{scope}
        \draw[rqcreamdivider,line width=0.55pt] (5.925,2.28) -- (5.925,8.55);
        \draw[rqcreamdivider,line width=0.55pt] (11.975,2.28) -- (11.975,8.55);

        \node[anchor=north west,inner sep=0pt] at (0.05,8.20) {%
            \includegraphics[width=5.65cm]{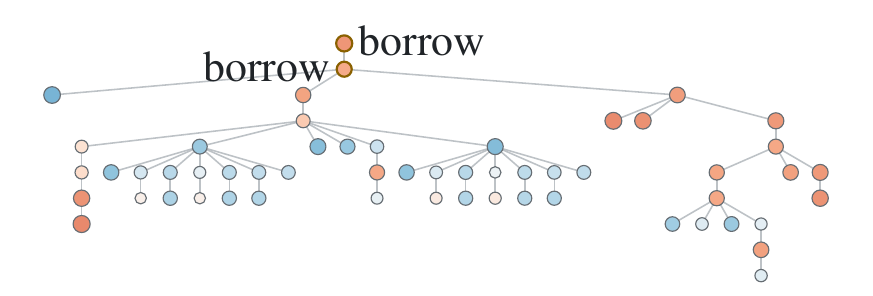}};
        \node[anchor=north west,inner sep=0pt] at (6.10,8.20) {%
            \includegraphics[width=5.65cm]{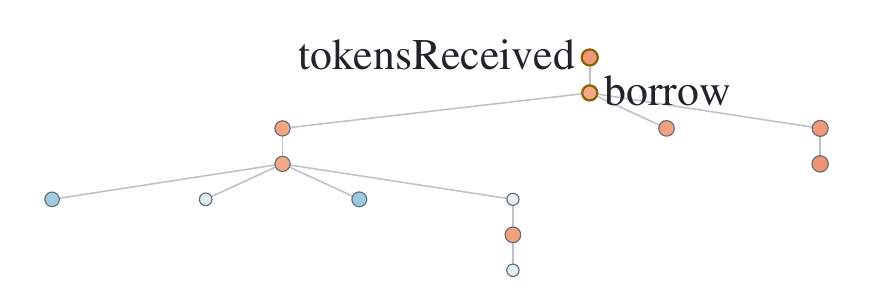}};
        \node[anchor=north west,inner sep=0pt] at (12.15,8.20) {%
            \includegraphics[width=5.65cm]{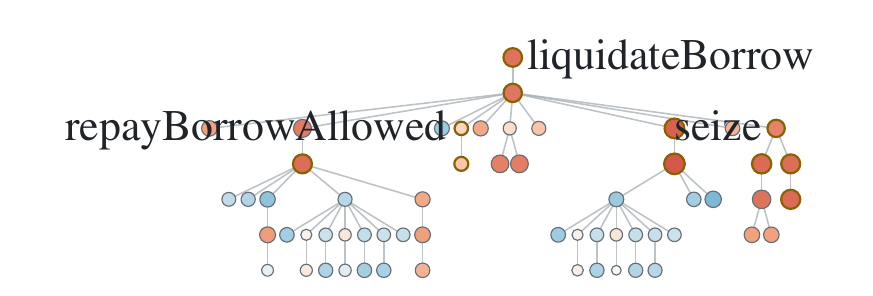}};

        \node[circle,fill=rqcreamgold,text=white,inner sep=1.8pt,font=\rmfamily\bfseries\footnotesize]
            at (0.22,8.38) {1};
        \node[rqtitle] at (0.56,8.38) {\textbf{Initial borrow} \(\cdot\) 56 calls};
        \node[circle,fill=rqcreamgold,text=white,inner sep=1.8pt,font=\rmfamily\bfseries\footnotesize]
            at (6.27,8.38) {2};
        \node[rqtitle] at (6.61,8.38) {\textbf{Reentrant borrow} \(\cdot\) 13 calls};
        \node[circle,fill=rqcreamgold,text=white,inner sep=1.8pt,font=\rmfamily\bfseries\footnotesize]
            at (12.32,8.38) {3};
        \node[rqtitle] at (12.66,8.38) {\textbf{Liquidation} \(\cdot\) 59 calls};

        \begin{scope}[yshift=1.20cm]
        \foreach \x in {0,6.05,12.10} {
            \node[anchor=north west,inner sep=0pt] at (\x+0.17,4.41) {%
                \includegraphics[width=5.41cm,trim=0 300bp 0 300bp,clip]
                    {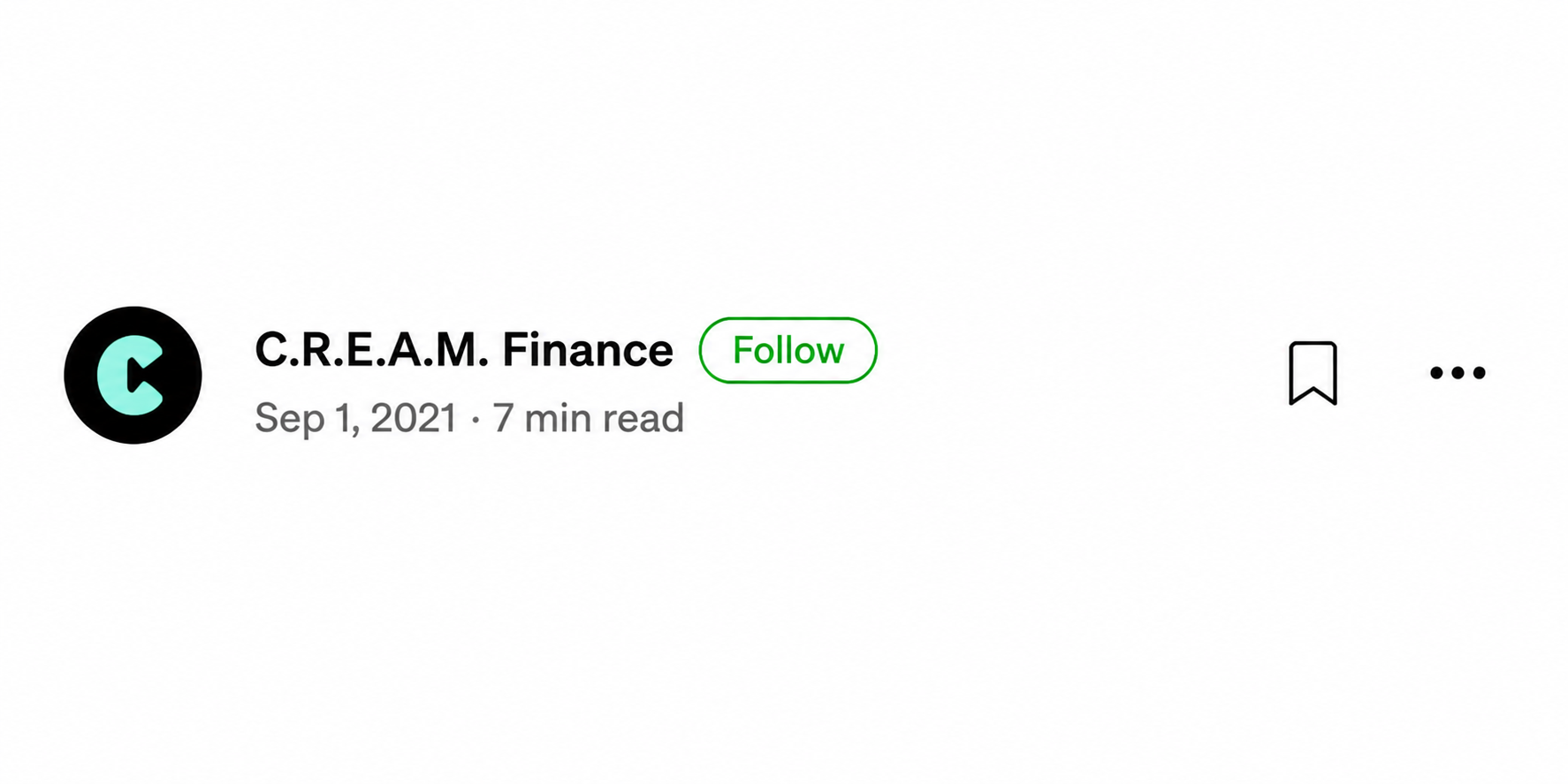}};
        }

        \node[rqcopy] at (0.27,3.15) {To be more specific, the vulnerability is};
        \node[rqcopy] (p1lead) at (0.27,2.72) {present within our};
        \node[rqcopy,rqpeach] (p1borrow) at ([xshift=0.07cm]p1lead.base east) {borrow function};
        \node[rqcopy] at ([xshift=0.07cm]p1borrow.base east) {when};
        \node[rqcopy] at (0.27,2.29) {executed in combination with \$AMP token's};
        \node[rqcopy] at (0.27,1.86) {ERC777 implementation, which calls the};
        \node[rqcopy] at (0.27,1.43) {tokenReceived hook.};

        \node[rqcopy] at (6.32,3.15) {Since \$AMP is an ERC777-like token,};
        \node[rqcopy,rqpeach] (p2token) at (6.32,2.72) {tokenReceived hook};
        \node[rqcopy] at ([xshift=0.07cm]p2token.base east) {is called.};
        \node[rqcopy] at (6.32,2.29) {This allowed the attacker to};
        \node[rqcopy,rqpeach] (p2borrow) at (6.32,1.86) {nest a second borrow() function};
        \node[rqcopy] at (6.32,1.43) {before the initial borrow()\ldots};

        \node[rqcopy] (p3lead) at (12.37,3.15) {Attacking contract B};
        \node[rqcopy,rqpeach] (p3liquidates) at ([xshift=0.07cm]p3lead.base east) {liquidates};
        \node[rqcopy] at ([xshift=0.07cm]p3liquidates.base east) {attacking};
        \node[rqcopy] (p3line2) at (12.37,2.72) {contract A by};
        \node[rqcopy,rqpeach] (p3repaying) at ([xshift=0.07cm]p3line2.base east) {repaying};
        \node[rqcopy] at ([xshift=0.07cm]p3repaying.base east) {38,960,000 AMP};
        \node[rqcopy] (p3line3) at (12.37,2.29) {back to crAMP market, and};
        \node[rqcopy,rqpeach] (p3seized) at ([xshift=0.07cm]p3line3.base east) {seized};
        \node[rqcopy] at (12.37,1.86) {35,529.572 crETH from attacking contract};
        \node[rqcopy] at (12.37,1.43) {A's collateral.};
        \end{scope}

        \draw[rqmap] (2.277,7.862) -- (p1borrow.north);

        \draw[rqmap] (9.922,7.773) -- (p2token.north);
        \draw[rqmap] (9.918,7.544) -- (p2borrow.north);

        \draw[rqmap] (15.471,7.749) -- (p3liquidates.north);
        \draw[rqmap] (15.138,7.309) -- (p3repaying.north);
        \draw[rqmap] (17.178,7.294) -- (p3seized.north);

        \foreach \panelx in {0,6.05,12.10} {
        \begin{scope}[shift={(\panelx,0)}]
            \node[rqlegendlabel,anchor=south west] at (0.13,6.23) {Benign};
            \node[rqlegendlabel,anchor=south east] at (5.67,6.23) {Attack};
            \shade[shading=rqcreamheatmap]
                (0.13,6.06) rectangle (5.67,6.20);
            \draw[rqcreamnodeborder,line width=0.25pt]
                (0.13,6.06) rectangle (5.67,6.20);
            \foreach \x in {0.572,1.458,2.577,2.900,3.223,4.342,5.228} {
                \draw[rqcreamnodeborder,line width=0.25pt] (\x,6.06) -- (\x,6.03);
            }
            \node[rqlegendtick] at (0.572,6.02) {\(-10^{-1}\)};
            \node[rqlegendtick] at (1.458,6.02) {\(-10^{-2}\)};
            \node[rqlegendtick] at (2.577,6.02) {\(-10^{-3}\)};
            \node[rqlegendtick] at (2.900,6.02) {\(0\)};
            \node[rqlegendtick] at (3.223,6.02) {\(10^{-3}\)};
            \node[rqlegendtick] at (4.342,6.02) {\(10^{-2}\)};
            \node[rqlegendtick] at (5.228,6.02) {\(10^{-1}\)};
        \end{scope}
        }
    \end{tikzpicture}%
    }
    \caption{Official incident report excerpts~\cite{cream-finance-blog} (bottom) aligned with three call trace subtrees containing positively attributed calls (top) for the \cream{}. Highlighted terms are exploit mechanisms. Blue/red denote negative/positive attribution. Connected lines indicate alignment between the documented exploit mechanism and the positively attributed call trace.}
    \label{fig:rq3-cream-explainability}
    \end{minipage}
\end{figure*}

%% file: assets/table-real-world-deployment-blockchain-throughput.tex
\begin{table}[!t]
    \centering
    \caption{Block time (s), transaction throughput per second (TPS), and transaction time (ms) of the seven most vulnerable blockchains in September 2025.}
    \label{tab:real-world-deployment-blockchain-throughput}
    \footnotesize
    \begin{tabular}{lrrrr}
        \toprule
        \textbf{Blockchain} & \textbf{Total Txn} & \textbf{Block (s)} & \textbf{TPS} & \textbf{Txn (ms)} \\
        \midrule
        Ethereum & 47,140,947 & 12.07 & 18.18 & 55.00 \\
        BSC & 418,450,657 & 0.75 & \underline{161.43} & \underline{6.19} \\
        Base & 350,258,844 & 2.00 & 135.13 & 7.40 \\
        Arbitrum & 84,269,190 & \underline{0.25} & 32.51 & 30.75 \\
        Avalanche & 49,406,364 & 1.61 & 19.06 & 52.46 \\
        Polygon & 108,646,847 & 2.15 & 41.92 & 23.86 \\
        Optimism & 36,105,960 & 2.00 & 13.92 & 71.82 \\
        \bottomrule
    \end{tabular}
\end{table}

%% file: assets/fig-rq5-feature-ablation.tex
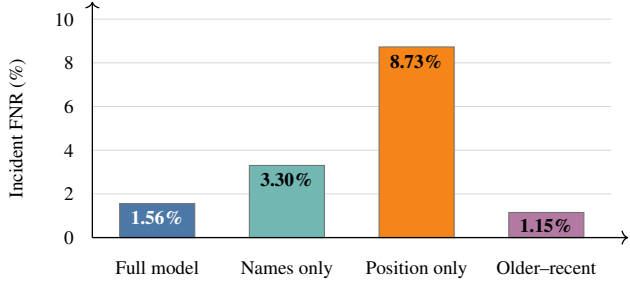
\begin{figure}[!t]
    \centering
    \definecolor{rqablblue}{HTML}{4C78A8}
    \definecolor{rqablteal}{HTML}{72B7B2}
    \definecolor{rqablorange}{HTML}{F58518}
    \definecolor{rqablpurple}{HTML}{B279A2}
    \definecolor{rqablgrid}{HTML}{D7D7D7}
    \resizebox{0.98\columnwidth}{!}{%
    \begin{tikzpicture}[x=1cm,y=0.32cm,font=\footnotesize]
        \foreach \y in {0,2,4,6,8,10} {
            \draw[rqablgrid,line width=0.35pt] (0,\y) -- (7.6,\y);
            \node[anchor=east] at (-0.15,\y) {\y};
        }

        \filldraw[fill=rqablblue,draw=black!55,line width=0.45pt]
            (0.4,0) rectangle (1.5,1.557);
        \filldraw[fill=rqablteal,draw=black!55,line width=0.45pt]
            (2.3,0) rectangle (3.4,3.302);
        \filldraw[fill=rqablorange,draw=black!55,line width=0.45pt]
            (4.2,0) rectangle (5.3,8.726);
        \filldraw[fill=rqablpurple,draw=black!55,line width=0.45pt]
            (6.1,0) rectangle (7.2,1.149);

        \node[anchor=north,fill=rqablblue,text=white,inner sep=0.8pt,
              font=\footnotesize\bfseries]
            at (0.95,1.35) {1.56\%};
        \node[anchor=north,fill=rqablteal,text=black,inner sep=0.8pt,
              font=\footnotesize\bfseries]
            at (2.85,3.10) {3.30\%};
        \node[anchor=north,fill=rqablorange,text=black,inner sep=0.8pt,
              font=\footnotesize\bfseries]
            at (4.75,8.50) {8.73\%};
        \node[anchor=north,fill=rqablpurple,text=black,inner sep=0.8pt,
              font=\footnotesize\bfseries]
            at (6.65,0.95) {1.15\%};

        \draw[->,line width=0.55pt] (0,0) -- (7.85,0);
        \draw[->,line width=0.55pt] (0,0) -- (0,10.8);
        \node[rotate=90,anchor=south,font=\footnotesize]
            at (-0.85,5) {Incident FNR (\%)};
        \node[anchor=north,align=center] at (0.95,-0.7) {Full model};
        \node[anchor=north,align=center] at (2.85,-0.7)
            {Names only};
        \node[anchor=north,align=center] at (4.75,-0.7)
            {Position only};
        \node[anchor=north,align=center] at (6.65,-0.7)
            {Older--recent};
    \end{tikzpicture}%
    }
    \caption{FNR comparison with feature and temporal ablation results. Lower is better.}
    \label{fig:rq5-feature-ablation}
\end{figure}

%% file: assets/list-call-trace-tree-min.tex
\begin{figure}[!t]
\centering
\begin{codebox}
\begin{lstlisting}[escapeinside={§}{§}]
§\textbf{[A]} 1: Attacker\arrow AC.start§
	§\textbf{[C]} 17: AC\arrow AMPMarket.borrow§
		§57: AMPMarket\arrow AMPToken.transfer§
			§60: AMPToken\arrow AC.tokensReceived§
				§\textbf{[C]} 61: AC\arrow ETHMarket.borrow§
					§70: ETHMarket\arrow AC.ether\_transfer§
	§74: AC\arrow AMPToken.transfer§
	§\textbf{[A]} 77: AC\arrow AC2.0x794404d§
		§\textbf{[C]} 79: AC2\arrow AMPMarket.liquidate§
			§131: AMPMarket\arrow ETHMarket.seize§
				§134: ETHMarket\arrow AC2.ether\_transfer§
		§\textbf{[C]} 139: AC2\arrow ETHMarket.redeem§
\end{lstlisting}
\end{codebox}
\captionsetup{name=Listing}
\caption{Minimal call traces for the \cream{}; [A] marks
attacker-controlled calls, and [C] marks critical ones.}
\label{lst:call-trace-tree-min}
\end{figure}

%% file: sections/07-discussion.tex
\section{Threats to Validity}
\label{sec:discussion}

\textbf{Internal Threats.}
\emph{Dataset correctness and representativeness.}
The DeFiHackLab dataset covers diverse blockchain application attacks (Section~\ref{sec:setup}), but its community-contributed labels may be inconsistent, affecting \sysname{} and other supervised detectors.
Prior studies instead curate datasets~\cite{wu2024dappfl,wu2025hunting,qin2025enhancing}, but these contain at most 200 transactions versus our 14,611 transactions.
For FPR, we assume the benign set contains no known attacks and cross-check it with our attack dataset, yet unknown attacks may remain.
For example, our web searching found one unreported attack not included by the DeFiHackLab repository, which was counted as a false positive.
Thus, the reported 0.0017\% FPR is a conservative upper bound: flagged transactions may be undiscovered attacks, making the true FPR lower.

\emph{Attacker attribution.}
RQ1 and RQ6 use the root call's callee as the proxy for the attacker contract, and treat other calls as victim-side.
Because attackers may use auxiliary contracts beyond that callee, the reported attacker call share therefore can be higher than the actual attacker call share.

\emph{Attack categorization.}
Our categorization in Appendix~\ref{sec:appendix-attack-categories} is derived from the community categorization of incidents in the DeFiHackLab repository~\cite{DeFiHackLabs}.
We also manually refine and merge some categories based on our understanding of blockchain application attacks, \eg both ``lack of access control'' and ``insufficient permission check'' go to our ``access control'' category.
Other tools categorize differently: Hunting~\cite{wu2025hunting} treats ``flash loan attack'' separately, whereas we treat flash loans as tools facilitating other attacks.
Such differences may affect the comparison results.
Appendix~\ref{sec:appendix-baseline-comparison} compares these tools on their datasets, partially mitigating this threat.

\textbf{External Threats.}
\emph{Data drift in blockchain application attacks.}
As DeFi evolves, new dApp logic and attack vectors may emerge while existing attacks become less relevant, requiring retraining after distribution shifts and adding maintenance overhead.
Reliance on transaction semantics rather than specific code patterns may nevertheless adapt better to evolving vectors and depend less on hand-coded patterns than rule-based approaches.
Indeed, Section~\ref{sec:RQ5} and Appendix~\ref{sec:appendix-baseline-comparison} show zero FNR on the pre-2022 CLUE and Hunting datasets and 1.15\% FNR in the older-to-recent ablation.
Data drift also affects the prior detectors~\cite{wang2022defiscanner,su2021evil,wu2023know,zhang2020txspector,wu2023defiranger,chen2024flashsyn,kong2023defitainter,xie2024defort}.

\emph{Private transaction submission.}
Pre-commit detection assumes an attack is public in the mempool or submitted to a participating builder or validator.
An attacker can instead use a private relay or bribe a non-participant to include it without public propagation; an external \sysname{} deployment then detects the attack only after its block becomes public.
This limitation is common to all transaction-based detectors. %

%% file: sections/08-related-work.tex
\section{Related Work}
\label{sec:related}

Transaction-level blockchain analysis uses representations at different granularities.
Account-transaction networks support clustering, deanonymization, spam, and money-laundering analysis, but omit internal contract calls~\cite{wu2021analysis,ron2013quantitative,maesa2017detecting,moreno2016listening,hu2019characterizing}.
DEFIER instead clusters internal call trace graphs to reconstruct multi-stage dApp attacks~\cite{su2021evil}.
Transfer-oriented detectors such as DeFiRanger, POMABuster, and DeFiScanner model token flows or events for particular behaviors~\cite{wu2023defiranger,xi2024pomabuster,wang2022defiscanner}.
TxSpector detects specified attack patterns from transaction traces, whereas MoE provides runtime verification over semantically lifted events~\cite{zhang2020txspector,xu2025quantitative}.
Both of them remain specialized to particular attack categories.

Sequence-based behavioral detection also has a long history outside blockchains.
Early host-based intrusion detectors characterized normal program behavior using short sequences of operating-system calls~\cite{forrest1996sense,hofmeyr1998intrusion}, and later malware classifiers learned from system-call sequences with neural models~\cite{kolosnjaji2016deep}.
These systems are qualitatively different both in terms of their features, and the latency constraints of blockchains; therefore, they are not applicable to the blockchain attack detection.

The closest application-attack detectors are CLUE, Hunting, and HOUSTON.
CLUE mines handcrafted patterns from execution property graphs; Hunting learns from transaction graphs, contract source, and comments; and HOUSTON builds per-dApp interaction and source-derived invariant models~\cite{qin2025enhancing,wu2025hunting,meng2026houston}.
These choices provide richer protocol-specific features but require graph reconstruction, source artifacts, historical transactions, or per-contract setup.
\sysname{} instead uses selector-derived name sequences and structural position, requiring neither source code nor per-dApp rules. We already compared quantitatively against CLUE and Hunting; 
see Appendix~\ref{sec:appendix-houston-interaction} for the HOUSTON comparison.

Adjacent systems synthesize exploits, counterattacks, or defensive mitigation transactions rather than perform general transaction detection~\cite{qin2023blockchain,zhang2023your,chen2024flashsyn,shou2024backrunner,zhou2021just}.
Trace-interpretation tools explain transactions but do not detect any attack~\cite{blocksec-metasuites,tenderly,blocksec}.
BlockGPT~\cite{gai2023blockchain} is an anomaly ranker, but reports per-contract anomaly rankings rather than binary predictions, so its results are not comparable in our evaluation.

%% file: sections/09-conclusion.tex
\section{Conclusion}
\label{sec:conclusion}

We propose \sysname{}, a blockchain application attack detector that leverages semantic information in transaction call traces. \sysname{} addresses the limitations of previous transaction analyzers that require human labour to craft patterns from reconstructed execution property graphs or require contract source code.
\sysname{} maps call traces to function name sequences to capture high-level semantics from a transaction, and learns transaction semantics from these sequences via a Transformer model for efficient attack detection.

Our empirical study shows that function name feature covers 98.46\% of calls, 23.68 percentage points higher than source code coverage.
\sysname{} achieves a 1.56\% incident-level FNR and a 0.0017\% transaction-level FPR, outperforming the length heuristic (31.31\% FNR), CLUE (70.77\%), and Hunting (83.17\%).
The detection results from \sysname{} are explainable.
We compare our detection results with the 184 corresponding available incident reports, and find a median alignment score of 74.86\%.
We also simulate possible attacker evasion scenarios, finding that the FNR increases only from 1.56\% to 2.26\%, demonstrating \sysname{}'s limited sensitivity to evasion.
Finally, its 24.90-millisecond mean detection time supports low-latency detection in popular blockchains.

%% file: appendices/01-ethics.tex
\section*{Ethical Considerations}
\phantomsection
\label{sec:appendix-ethics}

Our work focuses on detecting and analyzing attack transactions that have already been committed to public blockchains.
We neither conducted active attacks nor tested exploits against live blockchains.
During our research, we identified several issues in the DeFiHackLab repository~\cite{DeFiHackLabs} and in the Hunting~\cite{wu2025hunting} and CLUE~\cite{qin2025enhancing} datasets and implementations.
We responsibly disclosed these issues to the relevant maintainers and submitted corrective pull requests, all of which have since been merged into the corresponding repositories.
We omit nonessential communication details that could expose third-party identities.

%% file: appendices/02-open-science.tex
\section*{Open Science}
\phantomsection
\label{sec:appendix-open-science}

We will publicly release the datasets, source code, and scripts needed to reproduce the results and figures reported in this paper upon acceptance for publication.
The release will include metadata for all 688 incidents listed in the March 27, 2026 DeFiHackLab repository snapshot, including each incident's name, date, and URL, and will identify the 424 in-scope incidents used for training and evaluation.
It will include the 14,611 attack transactions (attack dataset) and the 134,665 victim historical transactions (victim dataset).
Because the full Ethereum benign dataset contains 537,069,442 transactions and is too large to distribute in the package, the release will include a subset of 171,584 transactions.
It will also include the selector-to-name database constructed from the Sourcify dataset to reproduce the RQ1 results, the source code for training \sysname{}, and the scripts used to calculate the FNR, FPR, and runtime overhead in RQ2.
For the FP analysis in RQ2, the release will provide records from exact-hash searches for all 9,228 transactions classified as positive.
For RQ3, it will further include the 184 incident reports and the script needed to reproduce the results.
We cannot, however, guarantee exact reproduction of the runtime results reported in RQ4 on hardware configurations that differ from ours, such as CPU-only servers.
Finally, the release will provide the scripts used to carry out the ablation and evasion studies and to generate the figures in RQ5 and RQ6.

%% file: assets/table-attack-categories.tex
\begin{table*}[!t]
    \centering
    \caption{Application-level attack categories considered in this work.}
    \label{tab:attack-categories}
    \footnotesize
    \begin{tabular}{@{}p{0.18\textwidth} p{0.76\textwidth}@{}}
        \toprule
        \textbf{Category} & \textbf{Explanation} \\
        \midrule
        Access control & Missing or flawed permission checks allow unauthorized execution of privileged functions (\eg emergency withdrawal, pool creation). \\
        Business logic & Violations of intended economic or functional invariants (\eg broken atomicity, collateralization design); see the example in Section~\ref{sec:background:motivating}. \\
        Price oracle manipulation (POMA)~\cite{xi2024pomabuster} & Distorting oracle-fed prices (often via DEX trades) to create artificial price divergence and profitable imbalances. \\
        Reentrancy~\cite{dhillon2017dao} & Re-entering a function before state updates finalize, often combined with other logic flaws to craft complex exploits. \\
        Token-specific & Exploiting non-standard or edge-case token behaviors (\eg inflationary supply) that the victim dApp fails to anticipate. \\
        Arbitrary call & Abusing call primitives to trigger unintended external calls in the victim contract. \\
        \bottomrule
    \end{tabular}
\end{table*}

%% file: appendices/03-solidity-evm-functions.tex
\section{How Solidity and EVM Handle Functions}
\label{sec:appendix:background:function}

We summarize how Solidity function signatures map to EVM execution and appear in transaction call traces.

\label{sec:background:function}

In Solidity, a function is defined by its signature, which consists of the name and parameter types, as shown in Listing~\ref{listing:cream-finance-code}.
After compilation, the function signature is hashed using Keccak-256, and the first 4 bytes of this hash become the function selector.
For example, selector \texttt{0xa9059cbb} corresponds to the ERC20 transfer signature \texttt{transfer(address,uint256)}.
This is analogous to a C program that maps compact identifiers to function implementations, for example through a \texttt{switch} statement or a function-pointer table, although C itself does not assign hashed identifiers to API functions~\cite{iso9899c18,kernighan1988c,harbison2002cref}.
The original analogy can be read as a three-level correspondence between Solidity/EVM execution and C execution.
At the definition level, a Solidity signature such as \texttt{transfer(address,uint256)} plays a role similar to a C interface such as \texttt{void transfer(void* addr,int val)}.
Both describe the callable operation and its expected arguments.
At the dispatch level, Solidity derives a 4-byte selector such as \texttt{0xa9059cbb} from the signature, whereas a C program could use an implementation-defined compact key, enum value, or function-pointer-table index to choose the target function.
At the tracing level, an EVM call trace records transaction-specific metadata, including sender, receiver, call type, selector, calldata, and return value. Analogously, a C tracing tool such as \texttt{uftrace} records caller--callee relationships and function-entry or function-exit events.
A transaction interacting with a dApp may span many call traces, depending on how many function calls it makes.

When a transaction calls a function, the EVM uses the function selector to determine which function to execute and generates a call trace.
A call trace also records the sender, receiver, call type, selector, calldata (parameters), and return value.
Analogously, a C function-call tracer such as \texttt{uftrace} can record a program's sequence of function entries and exits (caller and callee), yielding a call trace~\cite{uftrace,kernighan1999practice}.

Geth's \texttt{callTracer} represents a transaction as a tree of nested call frames, with the top-level call at the root and sub-calls as children~\cite{geth-calltracer}.

\input{assets/list-call-trace-tree.tex}

While most function-level semantics come from developer-defined contract functions, call frames also record contract creation, \texttt{SELFDESTRUCT}, and ETH transfers~\cite{geth-calltracer}, which contribute to transaction semantics and attack detection.
\textbf{Contract creation} records the deployment of a smart contract.
Failure to map these traces can cause detectors to miss attacks that depend on the deployment of new smart contracts.
\textbf{Contract suicide} records the use of the \texttt{SELFDESTRUCT} opcode, which leads to the removal of a contract from the blockchain and the transfer of any remaining ETH.
Failure to map these traces can cause detectors to miss value-transfer or contract-removal events that are part of an attack transaction.
\textbf{ETH transfer} records ETH movements between accounts, often within complex contract interactions.
Failure to map them can leave value-transfer semantics incomplete, which may cause detectors to miss attacks whose profit or execution path depends on native ETH movement.
We hard-encode them as the names shown above.

%% file: assets/list-call-trace-tree.tex
\begin{figure}[!t]
\centering
\begin{codebox}
\begin{lstlisting}[escapeinside={§}{§}]
1: Attacker§\arrow§AC.start
	3: AC§\arrow§Uniswap.swap
		4: Uniswap§\arrow§WETH.transfer
		5: Uniswap§\arrow§AC.uniswapV2Call
			7: AC§\arrow§WETH.withdraw
				8: WETH§\arrow§AC.ether_transfer
			9: AC§\arrow§ETHMarket.supply
			17: AC§\arrow§AMPMarket.borrow
				57: AMPMarket§\arrow§AMPToken.transfer
					60: AMPToken§\arrow§AC.tokensReceived
						61: AC§\arrow§ETHMarket.borrow
							70: ETHMarket§\arrow§AC.ether_transfer
			73: AC§\arrow§AC2.create_contract
			74: AC§\arrow§AMPToken.transfer
			77: AC§\arrow§AC2.0x794404d
				79: AC2§\arrow§AMPMarket.liquidate
					131: AMPMarket§\arrow§ETHMarket.seize
				139: AC2§\arrow§ETHMarket.redeem
				145: AC2§\arrow§WETH.deposit
				147: AC2§\arrow§WETH.transfer
				148: AC2§\arrow§AC.suicide_contract
			149: AC§\arrow§WETH.deposit
			151: AC§\arrow§WETH.transfer
	155: AC§\arrow§AMPToken.transfer
	159: AC§\arrow§WETH.transfer
\end{lstlisting}
\end{codebox}
\captionsetup{name=Listing}
\caption{Simplified call traces for the \cream{} attack transaction.}
\label{lst:call-trace-tree}
\end{figure}

%% file: assets/table-appendix-comparison.tex
\begin{table*}[!t]
\centering
\caption{FNR in cross-dataset evaluation. Parentheses include transactions a tool could
not process; bold marks each baseline on its source dataset; underlining
marks the lowest FNR in each row. The DeFiHackLab dataset row shows the incident count in parentheses and reports incident-level FNR.}
\label{tab:appendix-baseline-comparison}
\footnotesize
\begin{tabular}{@{}lcllrccc@{}}
\toprule
\textbf{Dataset} & \textbf{Source} & \textbf{Chain} & \textbf{Period} &
\textbf{\#Txn} & \textbf{CLUE} & \textbf{Hunting} & \textbf{\sysname{}} \\
\midrule
CLUE & \cite{zhou2023sok} & ETH & 2020-02--2022-04 & 139 &
\textbf{17.86\%} (17.86\%) & 95.68\% & \underline{0\%} \\
Hunting & \cite{wu2024dappfl} & ETH/BSC & 2020-09--2022-07 & 116 &
66.67\% (68.10\%) & \textbf{62.93\%} & \underline{0\%} \\
DeFiHackLab & \cite{DeFiHackLabs} & Various & 2017-07--2026-03 &
14,611 (424) & 70.77\% & 83.17\% &
\underline{1.56\%} \\
\bottomrule
\end{tabular}
\end{table*}

%% file: appendices/04-attack-categories.tex
\section{Taxonomy of Blockchain Application Attacks}
\label{sec:appendix-attack-categories}

This appendix summarizes the taxonomy of blockchain application attack categories used throughout the paper.

\textbf{Blockchain application attacks} occur when an attacker exploits application state or business logic within a dApp or across multiple dApps to produce an unauthorized or unintended outcome~\cite{zhou2023sok}.
Common attack vectors in blockchain application attacks include the exploitation of \emph{application-level vulnerabilities}, such as business-logic flaws and price-oracle manipulation across one or more dApps, as well as \emph{smart contract-level vulnerabilities}, such as access-control vulnerabilities and arbitrary calls within contract code~\cite{DeFiHackLabs}.
Table~\ref{tab:attack-categories} summarizes the attack categories used in our evaluation, based on prior work on blockchain application attacks~\cite{zhou2023sok, DeFiHackLabs}.

%% file: appendices/05-function-name-long-tail.tex
\section{\texorpdfstring{Long-Tail Distribution of Function Names}{Long-Tail Distribution of Function Names}}
\label{sec:appendix-function-name-long-tail}

We further examine how call frequency is distributed across the function names provided by the selector-to-name database in the RQ1 cohort. We count the frequency of each database-provided function name in the 5,632,984 selector-bearing call traces and rank the names by call frequency.
Recall from RQ1 that our selector-to-name database covers 98.46\% of these call traces.
We calculate end-to-end top-$k$ coverage as the number of calls resolved by the top-$k$ names divided by all 5,632,984 selector-bearing call traces, including those unresolved by the database.

\input{assets/fig-rq1-long-tail.tex}

\textbf{The function-name feature is concentrated in a vocabulary of only 1,000 names.} The top 10, 100, and 1,000 names cover 51.09\%, 82.83\%, and 97.26\% of call traces, respectively, while the remaining names form a long tail. Hence, a relatively small number of function names account for most calls in the observed call traces.

\textbf{Implication and future work.} This concentration suggests that even in settings where fewer selectors map to known function names, the most frequent names may still provide enough call-trace coverage to support effective \sysname{} detection. However, frequency coverage alone does not establish detection accuracy because rare names may carry disproportionate evidence for particular attacks. As future work, we plan to evaluate this hypothesis by systematically restricting the selector-to-name database to the top-$k$ most frequent names, mapping all other selectors to \texttt{[UNK\_SELECTOR]}, and measuring how \sysname{}'s detection accuracy changes as $k$ decreases.
Figure~\ref{fig:rq1-long-tail} visualizes this long-tail concentration.

%% file: assets/fig-rq1-long-tail.tex
\begin{figure}[!t]
\centering
\includegraphics[width=\columnwidth]{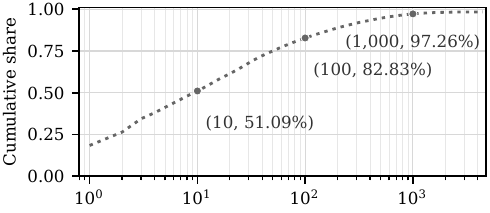}
\makebox[\columnwidth]{\footnotesize Function name counts, ordered by frequency}
\caption{Cumulative call coverage of database-provided function names in the combined repository cohort.}
\label{fig:rq1-long-tail}
\end{figure}

%% file: appendices/06-baseline.tex
\section{Experiments on Comparison Datasets}
\label{sec:appendix-baseline-comparison}

Section~\ref{sec:RQ2} reports an incident-level FNR of 1.56\% for \sysname{} on the attack dataset; Hunting and CLUE obtain incident-level FNR values of 83.17\% and 70.77\%, respectively, on the same set.
To better understand why Hunting and CLUE have high FNR values in our DeFiHackLab evaluation despite their reported effectiveness, and to support a fairer comparison, we evaluate \sysname{}, Hunting, and CLUE on both Hunting's and CLUE's own datasets.

Table~\ref{tab:appendix-baseline-comparison} gives an overview of the datasets. We obtain the CLUE dataset from the authors' GitHub repository~\cite{clue-repo} and the Hunting dataset from the DAppFL GitHub repository~\cite{wu2024dappfl}.
The CLUE dataset~\cite{qin2025enhancing} contains 140 attack transactions from two attack categories, Reentrancy and POMA, collected from Ethereum between February 17, 2020, and April 30, 2022. 
However, upon inspection, we find that one transaction is invalid, leaving
139 valid attack transactions in the dataset. We exclude that transaction
and will record its hash and exclusion reason in the artifact released upon acceptance.
The Hunting dataset~\cite{wu2025hunting} contains 116 attack transactions from DAppFL~\cite{wu2024dappfl} collected from Ethereum and BSC between September 13, 2020, and July 13, 2022.
It contains various attack categories, such as access control, POMA, business logic, and reentrancy.
We cross-check transactions from both CLUE and Hunting against our attack dataset of 14,611 attack transactions. 
Among 139 transactions in CLUE, 29 (20.86\%) occur in ours, while among 116 in Hunting, 24 transactions (20.69\%) do. 
Thus, neither comparison dataset is largely subsumed by our attack dataset.

\textbf{Compared with the complete DeFiHackLab dataset described in Section~\ref{sec:experiment}, both CLUE's and Hunting's datasets are much smaller.}
The CLUE dataset covers a narrower range of attack categories than DeFiHackLab and the Hunting dataset. 
Moreover, both the CLUE and Hunting datasets lack attack transactions from blockchains other than Ethereum and BSC, which limits the chain diversity of the datasets. 
Both the CLUE and Hunting datasets were collected before August 2022, while our DeFiHackLab dataset includes attack transactions through March 2026, which may introduce temporal bias in the comparison.

To understand the possible reasons behind the high FNR values reported by CLUE and Hunting, and to evaluate \sysname{}'s effectiveness under data drift, we re-evaluate all three detectors on both comparison datasets.
Table~\ref{tab:appendix-baseline-comparison} shows the results. 
The \textbf{bold} percentages mark each baseline evaluated on its own dataset, and the \underline{underlined} percentages indicate the lowest FNR for each dataset.
The numbers in parentheses for CLUE's FNR indicate the FNR including the transactions that CLUE fails to process.

\textbf{We observe that both Hunting and CLUE report lower FNR values on their own datasets}, with Hunting reporting a 62.93\% FNR on the Hunting dataset and CLUE reporting a 17.86\% FNR on the CLUE dataset.
This suggests that both Hunting and CLUE may have limited generalizability to other datasets.
These results also support the generalizability concerns discussed in Section~\ref{sec:RQ2}, including Hunting's dependence on historical contract-code and comment features and CLUE's fixed EPG patterns.

We communicated our configurations and results to the authors of the baseline methods.
Based on feedback from the authors, we identified a software dependency issue in CLUE and submitted a pull request correcting it.
We obtained our CLUE results using the corrected software version.
We also discussed Hunting's results with its authors.
They explained that their default configuration is precision-oriented and therefore prioritizes avoiding false positives over false negatives.
The authors also mentioned that tuning Hunting's parameters could lower its FNR; however, we decided to retain the default configuration reported in their paper.

\emph{\sysname{} achieves 0\% FNR on both the Hunting and CLUE datasets without retraining.}
These results show that function name sequences continue to generalize across the evaluated datasets, attack categories, and time periods.
The results support our design choice of modeling transaction semantics from function name sequences with a Transformer encoder, allowing \sysname{} to detect application-level attacks without hand-coded patterns or source code for analyzed contracts.
Together, the completed results provide evidence that \sysname{}'s transaction-semantics design contributes to its cross-dataset behavior, although differences in training-data scale and recency may also contribute.

%% file: appendices/07-houston.tex
\section{HOUSTON Interaction-Model Comparison}
\label{sec:appendix-houston-interaction}

Our initial baseline review excluded HOUSTON~\cite{meng2026houston} because its complete detector derives likely invariants from victim-contract source code. The public comparison datasets do not provide this source code, and the HOUSTON artifact does not supply the victim source required to run that pipeline.

On further inspection, we found that HOUSTON also contains an interaction model that learns historical protocol-interaction patterns without requiring contract source code. Because this component is independently runnable and produces transaction-level decisions comparable with \sysname{}, we evaluate interaction model here rather than treating it as the complete HOUSTON detector.
Table~\ref{tab:appendix-houston-interaction} summarizes this comparison.

\input{assets/table-appendix-houston.tex}

We reuse the HOUSTON attack dataset from the HOUSTON artifact, which contains 115 attack transaction hashes.
We exclude four attack transactions not available in our archive node and one transaction whose hash is invalid from its own dataset.
Our results show that HOUSTON's interaction model detects 82 of 110 evaluated attacks, corresponding to a 25.45\% FNR, and records an average runtime of 324.85 seconds.
In comparison, \sysname{} obtains a 2.61\% FNR on HOUSTON's dataset and an average runtime of 0.130 seconds.
\sysname{}'s runtime on HOUSTON's dataset is longer than on our attack and victim datasets because HOUSTON's dataset contains only attack transactions, which are more complex and time-consuming to analyze.

These results compare \sysname{} only with HOUSTON's interaction model on HOUSTON dataset. They do not reproduce or evaluate the complete source-code-dependent HOUSTON detector, and therefore are not included in the primary comparison in Section~\ref{sec:RQ2}.

%% file: assets/table-appendix-houston.tex
\begin{table}[!t]
    \centering
    \caption{Comparison with HOUSTON's runnable, source-code-independent interaction model.}
    \label{tab:appendix-houston-interaction}
    \footnotesize
    \begin{tabular}{lrr}
        \toprule
        \textbf{Method} & \textbf{FNR (\%)} & \textbf{Avg. time} \\
        \midrule
        HOUSTON interaction model & 25.45 & 324.85 s\\
        \sysname{} & 2.61  & 0.13 s\\
        \bottomrule
    \end{tabular}
\end{table}

%% file: appendices/08-false-positives-analysis.tex
\section{False Positives Analysis}
\label{sec:appendix-false-positives}

As mentioned in Section~\ref{sec:RQ2}, 
\sysname{} labels 9,228 transactions from the benign dataset as positive, but we do not understand
\begin{enumerate*}[label=(\arabic*)]
	\item whether any of these positives are real attacks that are not included in our attack dataset, and
	\item which function names attribute to the positive (attack) predictions.
\end{enumerate*}
Figure~\ref{fig:rq4-fp-categories} summarizes the function-name categories assigned to these predictions.
We therefore broaden our search on the 9,228 positives to find real attacks missing by our attack dataset, and then analyze the function name attribution of positive (attack) prediction and categorize them by type function names.
We conduct the analysis as follows:
\begin{enumerate*}[label=(\arabic*)]
	\item We search the web using the exact transaction hash of each FP and look for sources to identify the transaction as an attack. For transactions that match publicly reported attacks, we remove them from the FP analysis because they are known attacks rather than FPs.
	\item For the remaining transactions, we treat them as FPs. We compute LIG and rank function names by their positive attribution.
	\item We categorize FP transactions by the dominant function category.
\end{enumerate*}

\textbf{One apparent FP is a real attack.}
Our exact transaction hash search on Google identifies one of the 9,228 transactions as a real attack rather than an FP.
A contemporaneous report in GitHub links the exact Ethereum transaction hash to an exploit of improper access checks in the \texttt{ReferralFeeReceiver} contract with an approximately \$4.3K loss~\cite{franklin2024-improper-access}.
We exclude this transaction, leaving 9,227 FPs for the following analysis.
However, we cannot rule out that other FPs are real attacks that are not publicly reported or indexed by Google.
We are unable to manually verify the remaining 9,227 FPs due to the large amount of labour required.